\documentclass[aps,prl,twocolumn,superscriptaddress]{revtex4-2}
\usepackage{graphics,graphicx, array, afterpage}
\usepackage{qcircuit,amsmath}
\usepackage{grffile}
\usepackage[export]{adjustbox}
\usepackage{lineno}
\usepackage{xcolor}
\usepackage{braket}
\usepackage{array}
\usepackage{multirow}
\usepackage{enumerate}
\usepackage{amsmath}
\usepackage{amssymb}
\usepackage{amsthm}
\usepackage{times}
\usepackage{multirow}
\usepackage{natbib}
\usepackage[shortlabels]{enumitem}

\usepackage[colorlinks, citecolor=blue]{hyperref}
\hypersetup{linkcolor=blue,urlcolor=blue,citecolor=blue,pdfstartview={FitH},urlcolor=blue}

\DeclareMathOperator{\col}{col}
\newcommand{\hc}{\mathrm{h.c.}}
\renewcommand{\Re}{\mathop{\mathrm{Re}}}

\begin{document}

\title{Time Independence Does Not Limit Information Flow. I. The Free-Particle Case}


\author{Dong Yuan}
\affiliation{Center for Quantum Information, IIIS, Tsinghua University, Beijing 100084, People's Republic of China}
\affiliation{JILA, University of Colorado Boulder, Boulder, Colorado 80309, USA}

\author{Chao Yin}
\affiliation{Department of Physics and Center for Theory of Quantum Matter, University of Colorado Boulder, Boulder, Colorado 80309, USA}

\author{T. C. Mooney}
\affiliation{Joint Center for Quantum Information and Computer Science, NIST/University of Maryland, College Park, Maryland 20742, USA}
\affiliation{Joint Quantum Institute, NIST/University of Maryland, College Park, Maryland 20742, USA}

\author{Christopher L. Baldwin}
\affiliation{Department of Physics and Astronomy, Michigan State University, East Lansing, Michigan 48824, USA}

\author{Andrew M. Childs}
\affiliation{Joint Center for Quantum Information and Computer Science, NIST/University of Maryland, College Park, Maryland 20742, USA}
\affiliation{Department of Computer Science, University of Maryland, College Park, Maryland 20742, USA}
\affiliation{Institute for Advanced Computer Studies, University of Maryland, College Park, Maryland 20742, USA}

\author{Alexey V. Gorshkov}
\affiliation{Joint Center for Quantum Information and Computer Science, NIST/University of Maryland, College Park, Maryland 20742, USA}
\affiliation{Joint Quantum Institute, NIST/University of Maryland, College Park, Maryland 20742, USA}

\begin{abstract}
The speed of information propagation in long-range interacting quantum systems is limited by Lieb-Robinson-type bounds, whose tightness can be established by finding specific quantum state-transfer protocols.
Previous works have given quantum state-transfer protocols that saturate the corresponding Lieb-Robinson bounds using \textit{time-dependent} Hamiltonians. 
Are speed limits for quantum information propagation different for \textit{time-independent} 
Hamiltonians? In a step towards addressing this question, we present and analyze two optimal time-independent state-transfer protocols for free-particle systems, which utilize continuous-time single-particle quantum walks with hopping strength decaying as a power law.
We rigorously prove and numerically confirm that our protocols achieve quantum state transfer, with controllable error over an arbitrarily long distance in any spatial dimension, at the 
speed limits set by the free-particle Lieb-Robinson bounds. 
This shows that time independence does not limit information flow for long-range free-particle Hamiltonians.




\end{abstract}

\maketitle 

\textit{Introduction.}---The celebrated Lieb-Robinson bound~\cite{Lieb1972Finite} provides an upper limit on the speed of information propagation in non-relativistic quantum systems on lattices. 
Quantum state transfer~\cite{Bose2003Quantum,Yung2005Perfect,Burgarth2005Conclusive,Bose2007Quantum,Christandl2004Perfect,Albanese2004Mirror,Christandl2005Perfect,Nikolopoulos2004Coherent,Osborne2004Propagation,Li2005Quantum,Shi2005Quantum,Plenio2005High,Wojcik2005Unmodulated,Wojcik2007Multiuser,Fitzsimons2006Globally,Giovannetti2006Improved,Venuti2007Qubit,Franco2008Perfect,Markiewicz2009Perfect}
is a process through which an unknown quantum state on one site in a $d$-dimensional lattice is transferred to another site at a distance $L$. 
The two are closely related: Lieb-Robinson bounds limit the speed of quantum state transfer under unitary evolution, and conversely, finding specific state-transfer protocols that asymptotically achieve these speed limits establishes the tightness of Lieb-Robinson bounds.



Many quantum devices based on atomic, molecular, and optical systems naturally possess power-law interactions that decay as $1/r^\alpha$ with the distance $r$ for some exponent $\alpha$, including trapped ions~\cite{Monroe2021Programmable}, Rydberg atoms~\cite{Saffman2010Quantum}, nitrogen-vacancy centers~\cite{Doherty2013Nitrogen}, and polar molecules~\cite{Gadway2016Strongly}.
In these long-range interacting quantum systems, previous works have given state-transfer protocols~\cite{Guo2020Signaling,Tran2020Hierarchy,Tran2021Optimal} that use \textit{time-dependent} Hamiltonians to saturate Lieb-Robinson-type bounds~\cite{Chen2019Finite,Tran2019Locality,Kuwahara2020Strictly,Guo2020Signaling,Tran2021Lieb,Chen2023Speed}, establishing both the optimality of these protocols and the tightness of the bounds.
Since it can be easier to implement Hamiltonian dynamics without time-dependent control,
it is natural to ask whether the maximal speed of information flow is the same for time-independent long-range Hamiltonians. In other words, does the restriction of time independence slow down information propagation and lead to tighter Lieb-Robinson bounds? 

\begin{figure*}
\hspace*{-0.03\textwidth}
\includegraphics[width=0.97\linewidth]{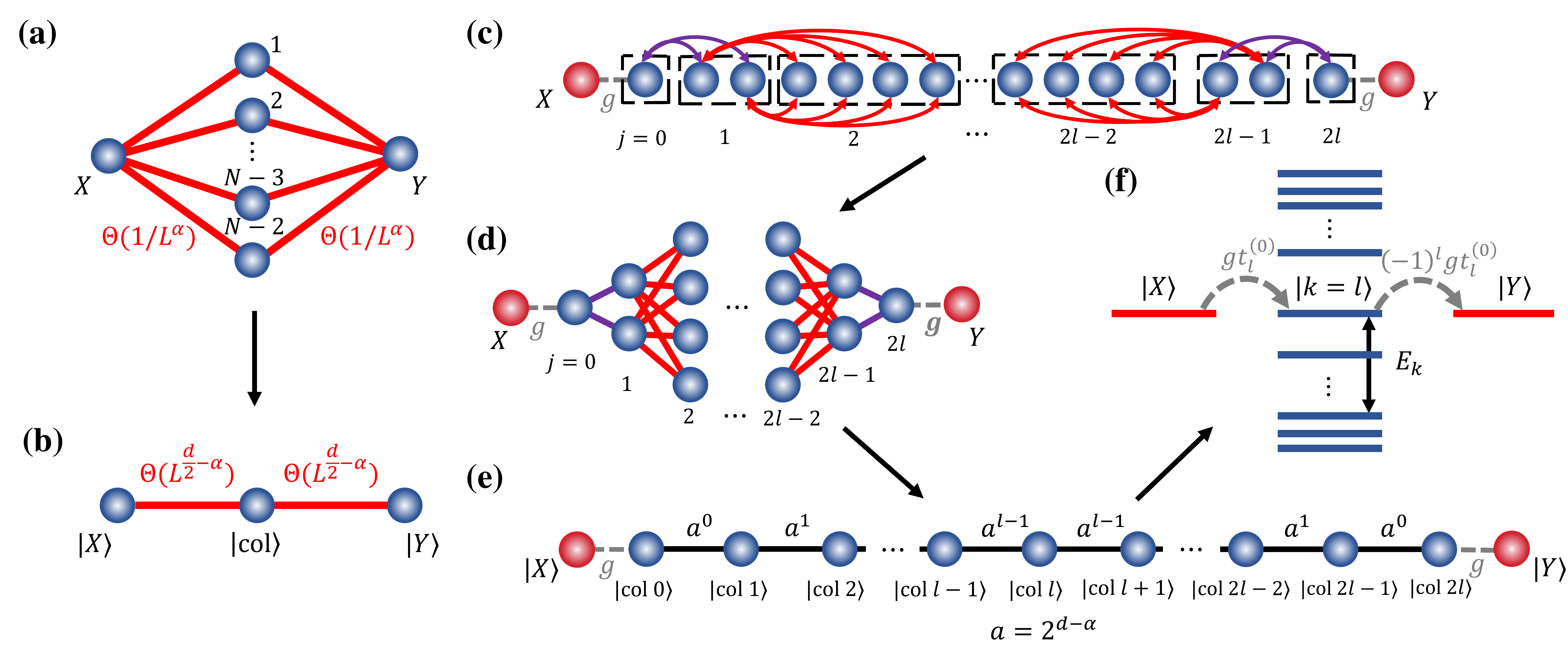} 
\caption{Summary of our two time-independent quantum state-transfer protocols via single-particle long-range quantum walks. (a)--(b) Illustrations of the optimal time-independent protocol for $\alpha<d/2$. The departure site $X$ and the arrival site $Y$ are connected through all other sites, with uniform hopping strengths $\Theta(1/L^\alpha)$ [(a)]. For the sites in the middle column, the only relevant state involved in the hopping process is the uniform superposition state $\ket{\col}$. Hence, the long-range hopping Hamiltonian reduces to a three-level Hamiltonian and admits perfect state transfer [(b)]. 
(c)--(f) Illustrations of the optimal time-independent protocol for $\alpha\ge d/2$. Instead of one-step transfer as in (a)--(b), we construct a multi-step recursive transfer protocol using a long-range quantum walk Hamiltonian [(c)]. By rearranging the sites [(d)] and identifying the relevant subspace during the hopping process (the uniform superposition state $\ket{\col\ j}$ of each column for $j=0$ to $2l$), we obtain a one-dimensional nearest-neighbor hopping graph with hopping strengths that exponentially increase (for $\alpha < d$) or decrease (for $\alpha > d$) away from the ends [(e)]. The single particle tunnels from (red) state $\ket{X}$ to (red) state $\ket{Y}$ through the zero-energy eigenstate $\ket{k=l}$ of the internal blue chain [(f)]. Other off-resonant eigenmodes lead to the state-transfer infidelity, which is made small by tuning the external coupling strength $g$. }
\label{fig:Illustration}
\end{figure*}

In this work, we provide a negative answer to the above question for the case of long-range free-particle Hamiltonians.
Specifically, we provide and analyze two time-independent state-transfer protocols illustrated in Fig.~\ref{fig:Illustration}, which are based on continuous-time single-particle quantum walks~\cite{Farhi1998Quantum,Childs2003Exponential,Childs2004Spatial,Childs2013Universal,Lewis2021Optimal} with hopping strength decaying as a power law.
We rigorously prove and numerically demonstrate that the state-transfer times  
of our protocols saturate the corresponding free-particle Lieb-Robinson bounds. Our results for the first time establish the optimality of time-independent state-transfer protocols using long-range Hamiltonians, and prove the tightness of long-range free-particle Lieb-Robinson bounds even when the restriction of time independence is imposed. 
In previous time-independent state-transfer protocols that employ long-range Hamiltonians~\cite{Avellino2006Quantum,Gualdi2008Perfect,Hermes2020Dimensionality,Lewis2023Ion}, 
the scaling behavior of the state-transfer times was not systematically analyzed. In contrast, our protocols are rigorously proved to have optimal quantum state-transfer times, and are able to transfer across an arbitrarily long distance $L$, in any spatial dimension $d$, for all power-law exponents $\alpha$, with an arbitrarily small constant error.





We consider bosonic and fermionic free-particle Hamiltonians with power-law long-range hopping on a $d$-dimensional cubic lattice. Such a system is governed by a Hamiltonian $H = \sum_{i \neq j} J_{i j} c^\dagger_i c_j + \sum_{i} \mu_i c^\dagger_i c_i $ with $|J_{ij}| \leq J_0 / r_{ij}^\alpha$ for some $\alpha \geq 0$. Here $c_i$ is the annihilation operator at site $i$ and $\mu_i$ is the chemical potential of site $i$. In other words, the absolute value of the hopping strength between two sites is upper bounded by a function that decays with their Euclidean distance $r_{ij}$ as a power law.
As proven in Refs.~\cite{Tran2020Hierarchy,Guo2020Signaling}, the operator norm of the (anti)commutator between the Heisenberg-picture-evolved operators $c_i(t)$ on site $i$ and $c_j^\dagger(0)$ on site $j$ (at distance $L$) is bounded above by
\begin{equation}
\| [c_i(t), c_j^\dagger(0)]_{\pm} \| \leq t \times 
\begin{cases}
\mathcal{O}( L^{d/2-\alpha}), & \alpha < d/2 \\
\mathcal{O}( 1 ), &  d/2 \leq \alpha < d \\
\mathcal{O}(L^{ d - \alpha}), &  d \leq \alpha < d+1 \\  
\mathcal{O}(L^{-1}), &  \alpha \ge d+1, 
   \end{cases}
\label{Eq:LR_bounds}
\end{equation}
where $[\cdot,\cdot]_{\pm}$ denotes the commutator for bosons ($+$) and the anticommutator for fermions ($-$).
The free-particle Lieb-Robinson bounds given by Eq.~\eqref{Eq:LR_bounds} hold for time-dependent $J_{ij}(t)$ and for arbitrarily strong time-dependent on-site terms $\mu_i(t)$~\cite{Tran2020Hierarchy,Guo2020Signaling}. 
Note that since they entail restrictions on the form of the Hamiltonian, free-particle Lieb-Robinson bounds establish tighter speed limits compared to the general interacting case with the same $\alpha$~\cite{Chen2019Finite,Tran2019Locality,Kuwahara2020Strictly,Tran2021Lieb}.



State transfer of a single particle from site $j$ to site $i$ with $\Omega(1)$ fidelity requires that $\| [c_i(t), c_j^\dagger(0)]_{\pm} \| = \Omega(1)$~\footnote{$\| [c_i(t), c_j^\dagger(0)]_{\pm} \| \ge |\bra{\mathrm{vac}} [c_i(t), c_j^\dagger(0)]_{\pm} \ket{\mathrm{vac}}| = |\bra{i} e^{-iHt} \ket{j}|$, where $\ket{\mathrm{vac}}$ denotes the single-particle vacuum state}. 
As a result, the above free-particle Lieb-Robinson bounds constrain the minimal time required for quantum state transfer. 
Previous works~\cite{Guo2020Signaling,Tran2020Hierarchy} have constructed two \textit{time-dependent} free-particle state-transfer protocols that saturate these bounds for all power-law exponents $\alpha$. In the next two sections, we ``staticize'' (i.e., make time-independent) the free-particle protocols of Ref.~\cite{Guo2020Signaling} and Ref.~\cite{Tran2020Hierarchy}, respectively, to realize optimal  \textit{time-independent} quantum state transfer. 
We emphasize that it is not clear whether any given protocol can be staticized without a slow down. Our approach to staticization depends on the setting---for example, our time-independent protocol for $\alpha \ge d/2$ relies on a delicate recursive construction of long-range Hamiltonians combined with a quantum tunneling trick.
Taken together, our protocols achieve state-transfer time $T=\mathcal{O}( L^{\alpha - d/2})$ for $\alpha < d/2$, $T=\mathcal{O}( 1 )$ for $d/2 \leq \alpha < d$, $T = \mathcal{O}(\log L)$ for $\alpha = d$, and $T = \mathcal{O}( L^{\alpha - d})$ for $d < \alpha < d+1$, which are optimal---i.e., saturate the bounds in Eq.~\eqref{Eq:LR_bounds}---up to subpolynomial corrections~\footnote{For $\alpha\ge d+1$, Eq.~\eqref{Eq:LR_bounds} can be easily saturated by nearest-neighbor hopping models \cite{Yao2011Robust}.}.   



\textit{The protocol for $\alpha < d / 2$.}---For  $\alpha < d/2$, since  particle hopping is strongly long-range so that every site is appreciably coupled to all other sites, we construct the following simple time-independent Hamiltonian to achieve optimal state transfer. In a $d$-dimensional cubic lattice with side-length $L$, we connect the departure site $X$ [with coordinates $(0,0,\ldots,0)$] and the arrival site $Y$ [with coordinates $(L-1,0,\ldots,0)$] with all the other sites [labeled as $1,2,\ldots, N-3, N-2$, where $N=L^d$].
As shown in Fig.~\ref{fig:Illustration}(a), crucially we take all the hopping strengths to be the same, equal to $1/(\sqrt{d}L)^\alpha = \Theta(1/L^\alpha)$. Therefore, for the sites in the middle column, the only relevant state involved in the hopping process is the uniform superposition state $\ket{\col} = \frac{1}{\sqrt{N-2}} \sum_{i=1}^{N-2} \ket{i}$ [Fig.~\ref{fig:Illustration}(b)]. This procedure is described by the Hamiltonian
\begin{align}
H & = \frac{1}{{(\sqrt{d}L)}^\alpha}\sum_{i=1}^{N-2} (\ket{X} \bra{i} + \ket{i} \bra{Y} + \hc )  \nonumber \\
& = \frac{\sqrt{N-2}}{{(\sqrt{d}L)}^\alpha}(\ket{X}\bra{\col}+\ket{\col}\bra{Y} + \hc). 
\end{align}
Hence, the long-range hopping Hamiltonian for $N$ sites reduces to the Hamiltonian of a system with the three levels $\{\ket{X}, \ket{\col}, \ket{Y}\}$, which can be solved exactly and admits perfect single-particle transfer from $\ket{X}$ to $\ket{Y}$ in time
\begin{equation}
T = \frac{\pi}{\sqrt{2}}\frac{{(\sqrt{d}L)}^\alpha}{\sqrt{N-2}} = \mathcal{O}(L^{\alpha - d/2}).
\label{Eq:T_smallalpha}
\end{equation}
This indeed saturates the free-particle Lieb-Robinson bound, Eq.~\eqref{Eq:LR_bounds}, for $\alpha < d/2$.

\textit{The protocol for $\alpha \ge d / 2$.}---When the power-law exponent increases to $\alpha \ge d / 2$, the one-step state-transfer protocol in Fig.~\ref{fig:Illustration}(a)--(b) is no longer optimal. Instead, we construct a multi-step recursive transfer protocol using a long-range quantum walk Hamiltonian. As shown with blue sites in Fig.~\ref{fig:Illustration}(c) for $d = 1$, we divide the lattice into multiple blocks (in higher dimensions we select a collection of blocks from the $d$-dimensional lattice). The $j$th block is a $d$-dimensional hypercube with side length $l_j = 2^{\min\{j, 2l - j\}}$ (the base $2$ can be replaced by any other integer larger than $1$), and the index $j$ takes integer values $0,1,2,\ldots,l-1,l,l+1,\ldots,2l-1,2l$. Neighboring blocks are disjoint and touch only at their corners [we present an illustration of the block arrangement for higher dimensions in Fig.~\ref{fig:Illustration_d=2} of the Supplemental Material (SM)]. The $j$th block contains $N_j = (l_j)^d$ sites, and the quantum state-transfer distance $L = \sum_{j=0}^{2l} l_j= 2^{l+1} + 2^{l} - 2$ increases exponentially with $l$.
Similarly to our protocol for $\alpha < d/2$, we take all hopping strengths between sites in the $j$th block and those in the $(j+1)$st block to be $\bigl[ \sqrt{ d \times (l_j + l_{j+1})^2 } \bigr] ^{-\alpha}$. The quantity in the square brackets corresponds to the maximum distance between sites in the $j$th and $(j+1)$st block. After neglecting an unimportant overall constant, the hopping strength is proportional to $(2^{\min\{j, 2l - 1 - j\}})^{-\alpha}$.





To better visualize our protocol, we rearrange the vertices and hopping edges to obtain the graph shown in Fig.~\ref{fig:Illustration}(d).
Since the hopping strengths of edges between each two neighboring columns are the same, the relevant subspace during the hopping process consists of the uniform superposition state $\ket{\col  j} = \frac{1}{\sqrt{N_j}} \sum_{i \in \text{column } j}\ket{i}$ of each column. Therefore, we reduce the $d$-dimensional power-law long-range hopping Hamiltonian into a one-dimensional nearest-neighbor quantum walk Hamiltonian with hopping strengths $a^0, a^1, a^2,\ldots, a^{l-2},a^{l-1},a^{l-1},a^{l-2},\ldots,a^1, a^0$ (up to an overall constant), where $a= 2^{d-\alpha}$ [Fig.~\ref{fig:Illustration}(e)].



Next, in order to \textit{time-independently} control the state-transfer infidelity, we adopt the quantum tunneling trick used in Refs.~\cite{Li2005Quantum,Plenio2005High,Wojcik2005Unmodulated,Wojcik2007Multiuser,Yao2011Robust,Yao2013Quantum}. Specifically, we attach the (red) departure site $X$ and the (red) arrival site $Y$ to the two end blocks, and utilize all the blue sites in Fig.~\ref{fig:Illustration}(c)--(e) as the state transfer channel. The parameter $g$ controls the coupling strength between the two external sites and the transfer channel. As shown in Fig.~\ref{fig:Illustration}(f), after we diagonalize the blue chain in Fig.~\ref{fig:Illustration}(e),  we obtain the Hamiltonian $H = \sum_{k=0}^{2 l} E_k |k\rangle\langle k|  + g \sum_{k=0}^{2 l}  t_k^{(0)} \left[ |X\rangle\langle k| + (-1)^k |k\rangle\langle Y| + \hc \right]$.
Here $k = 0,1,2,\ldots,2l$ labels the eigenstates from top to bottom of the spectrum.
As discussed in Section~\ref{sec:transfer_infidelity} of the SM, the single-particle spectrum $\{E_k\}$ of the blue chain is symmetric about zero, and there always exists a zero-energy eigenstate $\ket{k=l}$ that is resonant with the two external sites $X$ and $Y$. The coefficients $t_k^{(j)}$ correspond to the amplitudes of $\ket{\col j}=\sum_{k=0}^{2l}t_k^{(j)} \ket{k}$, with $t_k^{(j)} = (-1)^k\  t_k^{(2l-j)}$ due to the spatial inversion symmetry of the blue chain.




\begin{figure}[t!]
\hspace*{-0\textwidth}
\includegraphics[width=1\linewidth]{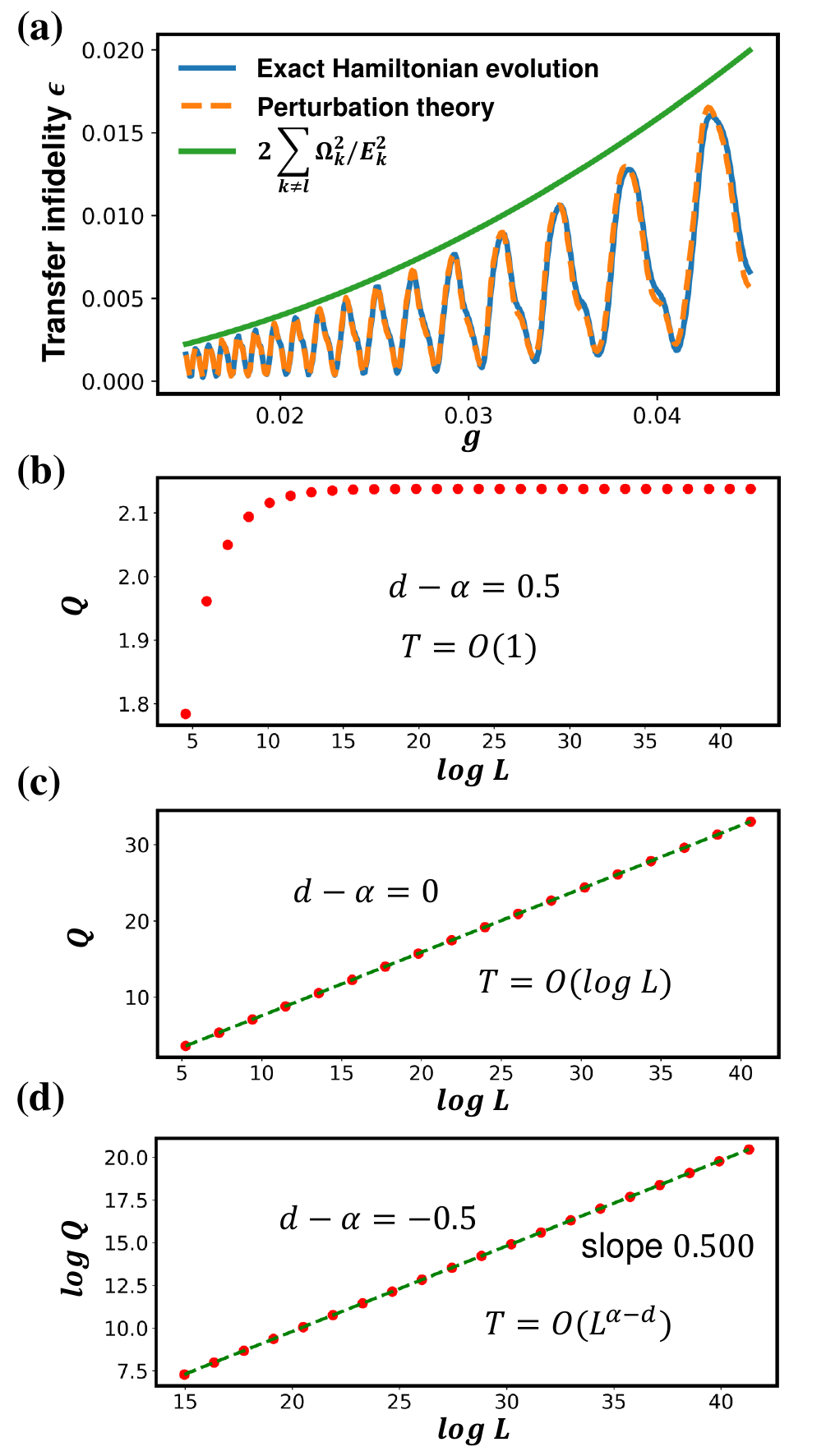} 
\caption{Numerical calculations of the state-transfer infidelity and the state-transfer time scaling of our time-independent protocol for $\alpha \geq d/2$. (a) State-transfer infidelity $\epsilon$ as a function of the coupling strength $g$. We take $d-\alpha=0.2$ and $l=24$ for the hopping graph in Fig.~\ref{fig:Illustration}(e). (b)--(d) The scaling behavior of the quantity $Q = \sqrt{\sum_{k\neq l} \bigl(\frac{t_k^{(0)}/t_l^{(0)}}{E_k}\bigr)^2}$ as a function of the state-transfer distance $L = 2^{l+1} + 2^{l} - 2$ for the three cases $\alpha < d $, $\alpha = d $, and $\alpha > d $, respectively. The red dots are numerical data points, and the green dashed lines show the results of linear fits. }
\label{fig:Numerics}
\end{figure}




When the coupling strength $g t_k^{(0)}$ is small enough compared to the detuning $E_k$ between the resonant and off-resonant levels, up to zeroth order in $g t_k^{(0)} / E_k$, the zero-energy eigenmode mediates tunneling from site $X$ to site $Y$ in time $T = \pi / (\sqrt{2} g t_l^{(0)}) = \pi / \Omega_l$ (where we define $\Omega_k = \sqrt{2} g t_k^{(0)}$), which is similar to the three-level result in Eq.~\eqref{Eq:T_smallalpha}. 
Beyond zeroth order, the coupling between the off-resonant levels and $\ket{X}, \ket{Y}$ makes the state transfer imperfect.
Standard perturbation theory~\cite{Yao2013Quantum} predicts that the state-transfer infidelity $\epsilon \equiv 1 - |\bra{Y} e^{-iHt} \ket{X}|^2$ is given to the leading order by (details presented in Section~\ref{sec:transfer_infidelity} of the SM)
\begin{equation}
\epsilon = \sum_{k \neq l} \Omega_k^2 [1 + (-1)^k \cos(E_k T)] / E_k^2.
\label{Eq:Infidelity_perturbation}
\end{equation}



To benchmark the accuracy of the perturbation theory calculations, we directly perform Hamiltonian evolution of the hopping model in Fig.~\ref{fig:Illustration}(e) from the initial state $\ket{X}$ for time $T = \pi / \Omega_l$ and compute the state fidelity with respect to $\ket{Y}$. As shown in Fig.~\ref{fig:Numerics}(a), the state-transfer infidelity after the Hamiltonian evolution is well captured by the perturbative result in Eq.~\eqref{Eq:Infidelity_perturbation}.

We observe from Fig.~\ref{fig:Numerics}(a) and Eq.~\eqref{Eq:Infidelity_perturbation} that, when we decrease the external coupling strength $g$, the leading-order transfer infidelity is always upper bounded by $2 \sum_{k \neq l} \Omega_k^2 / E_k^2$. 
Moreover, we rigorously prove an upper bound $\mathcal{O}( \sum_{k \neq l} \Omega_k^2 / E_k^2 )$ for the state-transfer infidelity in Section~\ref{sec:transfer_infidelity} of the SM.
Therefore, in order for the infidelity to be less than $\epsilon$, it suffices to set $g = \Theta(\epsilon^{1/2} [\sum_{k \neq l} (t_k^{(0)})^2/E_k^2]^{-1/2})$. 
Since the state transfer is achieved withi time $T = \mathcal{O}(1/g t_l^{(0)})$, we derive that
\begin{equation}
T  = \mathcal{O}\left[ \sqrt{ \frac{1}{\epsilon} \sum_{k \neq l} \left(\frac{t_k^{(0)}/t_l^{(0)}}{E_k}\right)^2 }\  \right].
\label{Eq:scaling_time_vs_eps_Delta}
\end{equation}


Thus the remaining task is to understand how the quantity $Q = \sqrt{\sum_{k \neq l} \left(\frac{t_k^{(0)}/t_l^{(0)}}{E_k}\right)^2} $ scales with the transfer distance $L$ for different $\alpha$ and $d$. This scaling is solely determined by the eigenvalues and eigenstates of the one-dimensional nearest-neighbor quantum walk Hamiltonian in Fig.~\ref{fig:Illustration}(e). 
As shown in Fig.~\ref{fig:Numerics}(b)--(d), we numerically calculate the quantity $Q$ for three cases: (1) When $\alpha < d $, $Q$ converges to a constant as $\log L$ increases, implying that $T=\mathcal{O}(1)$; (2) When $\alpha = d$, $Q$ is proportional to $\log L$, so $T = \mathcal{O}(\log L)$; (3) When $\alpha > d$,  $\log Q$ is proportional to $\log L$, and the slope is equal to $\alpha - d$, which implies that $T = \mathcal{O}(L^{\alpha - d})$. These scaling behaviors of the state-transfer times $T$ with respect to $L$, $\alpha$, and $d$ saturate the free-particle Lieb-Robinson bounds Eq.~\eqref{Eq:LR_bounds} (up to a logarithmic factor) and thus demonstrate the optimality of our time-independent protocol for $\alpha \ge d/2$.


We now 
prove the above scaling results for $Q$ and hence for the state-transfer time. For $\alpha = d$ (so $a=1$), the hopping Hamiltonian reduces to the one-dimensional nearest-neighbor quantum walk with uniform hopping strengths.
The eigenvalues and eigenstates of the Hamiltonian can be determined analytically:
$E_k = 2\cos \frac{(k+1)\pi}{2l+2}$ and $t_k^{(0)}/t_l^{(0)} = \sin \frac{(k+1)\pi}{2l+2}$. Hence, $\sum_{k \neq l} (\frac{t_k^{(0)}/t_l^{(0)}}{E_k})^2 = \mathcal{O}( l\times \int_{0}^{\frac{\pi}{2}-\frac{\pi}{2l+2}} \tan^2 x \, \mathrm{d}x ) = \mathcal{O}( l\times \cot \frac{\pi}{2l+2} ) = \mathcal{O}( l^2 )$, which directly gives $T = \mathcal{O}(l) = \mathcal{O}(\log L)$.
For $\alpha < d$ and $\alpha > d$ (i.e., $a > 1$ and $a < 1$, respectively), we can exactly determine the resonant zero-energy eigenstate $\ket{k=l}$: $t_{l}^{(2j)} = (-a)^{-j} \times \sqrt{a^{-l}+2(1-a^{-l})/(1-a^{-2})}$, $t_{l}^{(2 j+1)} = 0$. However, the eigenvalues and eigenstates of the off-resonant levels are no longer exactly solvable. Instead, we prove that $E_{l-1}$ (the smallest energy gap) is lower bounded by $\Omega(1)$ when $a>1$ and $\Omega(a^{l})$ when $a<1$.
The essential ideas of the proof are as follows (see Section~\ref{sec:proof_scaling} of the SM for details). We use the eigenstate equation $H \ket{k} = E_k \ket{k}$ to obtain recurrence relations for the amplitudes $\{t_k^{(j)}\}$: $E_k t_k^{(0)} = a^0 t_k^{(1)}$, $E_k t_k^{(1)} = a^0 t_k^{(0)} + a^1 t_k^{(2)}$, $\ldots$, $E_k t_k^{(l-1)} = a^{l-2} t_k^{(l-2)} + a^{l-1} t_k^{(l)}$, and $E_k t_k^{(l)} = a^{l-1} t_k^{(l-1)} + (-1)^k a^{l-1} t_k^{(l-1)}$. 
If there were an eigenstate with $E_k=o(1)$ for $a>1$ [or $E_k=o(a^l)$ for $a<1$] ($k\neq l$), the small parameter $E_k$ could be utilized to recursively upper and lower bound the coefficients $\{t_k^{(j)}\}$. We could then use these bounds to show that the last equality of the above recurrence relations [$E_k t_k^{(l)} = a^{l-1} t_k^{(l-1)} + (-1)^k a^{l-1} t_k^{(l-1)}$] is unsatisfiable, so there is no such eigenstate.






Next, we can upper bound the quantity $Q$ by exploiting the orthonormality of the coefficients $\{t_k^{(j)}\}$. For $a>1$, we have  $1/t_l^{(0)} = \mathcal{O}(1)$ and $E_k = \Omega(1)$, so
\begin{equation}
\sum_{k\neq l} \left[\frac{t_k^{(0)}/t_l^{(0)}}{E_k}\right]^2 \leq \text{const} \times \sum_{k\neq l} [ t_k^{(0)} ]^2 = \mathcal{O}(1).
\end{equation}




For $a<1$, since $1/t_l^{(0)} = \mathcal{O}(a^{-l/2})$, naively replacing all the $\{E_k\}$ by $\Omega(a^l)$ will not give the desired bound. We further observe from the recurrence relations of $\{t_k^{(j)}\}$ that 
$t_k^{(0)} = E_k [ t_k^{(1)} - t_k^{(3)} / a + t_k^{(5)} / a^2 + \cdots + t_k^{(l-1)} / (-a)^{l/2-1} ] + (-a)^{l/2} t_k^{(l)} $, and derive the following:
\begin{widetext}
\begin{align}
\sum_{k\neq l} \left[\frac{t_k^{(0)} / t_l^{(0)}}{E_k}\right]^2 & 
\leq 2a^{-l} \times \sum_{k\neq l} \left[ t_k^{(1)} - \frac{t_k^{(3)}}{a} + \frac{t_k^{(5)}}{a^2} + \cdots
+ \frac{t_k^{(l-1)}}{(-a)^{l/2-1}}  \right]^2 + \text{const} \times a^{-2l} \nonumber \\
& = 2a^{-l} \times (1 + \frac{1}{a^2} + \frac{1}{a^4} + \cdots + \frac{1}{a^{l-2}}) + \text{const} \times a^{-2l} = \mathcal{O}(a^{-2l}) = \mathcal{O}\left[L^{2(\alpha - d)} \right].      
\end{align}
\end{widetext}
For the first equality of the second line, the cross terms vanish since $\langle \col p |\col q \rangle = \sum_{k=0}^{2l} t_k^{(p)} t_k^{(q)} = \sum_{k\neq l} t_k^{(p)} t_k^{(q)} = \delta_{pq} $ for $p, q$ odd [note that $t_{l}^{(2 j+1)} = 0$]. 
Our rigorous proof confirms the state-transfer time scaling observed numerically, thereby saturating the free‐particle long-range Lieb-Robinson bounds.





\textit{Conclusion and outlook.}---We presented and analyzed two time-independent quantum state-transfer protocols that together saturate the long-range free-particle Lieb-Robinson bounds for all power-law exponents $\alpha$. Our study has both theoretical and practical implications.

Conceptually, we demonstrate, for the first time, the tightness of these fundamental Lieb-Robinson bounds in \textit{time-independent} long-range quantum systems, making Lieb-Robinson bounds directly relevant to systems without time-dependent control. Our results, especially the multi-step recursive construction, show how to achieve the fastest possible information propagation with continuous-time long-range quantum walks.





Practically, our time-independent state-transfer protocols have several desirable properties and imply optimal designs for future experiments and quantum computing architectures.
For instance, implementing our fully time-independent analog evolution simplifies the experimental control sequences.
Our protocols also enable a bus-based method to transfer quantum information in large-scale quantum computers (we call the resonant eigenmode through which the state transfers the ``bus mode"), which may provide a useful gadget for more complicated circuit operations, such as quantum routing~\cite{Bapat2023Advantages}. In Section~\ref{sec:another_protocl} of the SM, we adopt similar techniques to analyze 
another time-independent long-range free-particle protocol with hopping strengths exactly following $J_{ij} = J_0 / r_{ij}^\alpha$. This protocol can be naturally realized in atomic and molecular platforms. 
We demonstrate, for example, for one-dimensional trapped-ion crystals with $\alpha\approx 1$~\cite{Richerme2014Nonlocal,Jurcevic2014Quasiparticle}, this time-independent protocol can achieve state transfer in time $T=\mathcal{O}(\sqrt{L})$. Despite being suboptimal, it still provides a quadratic speed-up over a short-range protocol, showing the potential of using time-independent long-range Hamiltonians to accelerate quantum information processing tasks.






Compared with previous state-transfer protocols, our time-independent protocols are as fast as the time-dependent interacting protocol in Ref.~\cite{Eldredge2017Fast}, yet are slower than the optimal time-dependent interacting protocols in Refs.~\cite{Tran2021Optimal}. However, our free-particle protocols utilize $W$ states as the intermediate states (which offer robustness against errors in the Hamiltonian, as analyzed in Refs.~\cite{Tran2020Hierarchy,Hong2021Fast}), rather than going through the fragile Greenberger–Horne–Zeilinger states as in Refs.~\cite{Eldredge2017Fast,Tran2021Optimal,Yin2025Fast}. 
Moreover, similarly to the short-range case~\cite{Yao2011Robust}, the quantum tunneling process in our protocols actually implements a SWAP gate between the departure and arrival sites. Therefore, our time-independent free-particle state-transfer protocols do not depend on whether the intermediate sites are occupied, providing so-called unconditional quantum state transfer.


In our companion paper~\cite{Mooney2025Time}, we utilize a modified Kitaev clock construction~\cite{Kitaev2002Classical,Kempe2006Complexity,Watkins2024Time} to staticize the optimal state-transfer protocols using 
long-range interacting Hamiltonians~\cite{Tran2021Optimal}, by introducing ancilla qubits as clock registers while preserving the long-range locality.
We hope these investigations will motivate further exploration of speed limits for information flow in long-range interacting quantum systems with time-independence constraints, or more generally, with other conserved quantities or physical constraints.


\begin{acknowledgments}
We thank Dong-Ling Deng, Dhruv Devulapalli, Luming Duan, Adam Ehrenberg, Weiyuan Gong, Zhe-Xuan Gong, Liang Jiang,  Andrew Lucas, and Qi Ye for helpful discussions. D.Y.~acknowledges support from the National Natural Science Foundation of China (Grants No.~123B2072). C.Y.~was supported by the Department of Energy under Quantum Pathfinder Grant DE-SC0024324.
C.L.B.~was supported by start-up funds from Michigan State University.
T.C.M., A.M.C., and A.V.G.~were supported in part by the DoE ASCR Quantum Testbed Pathfinder program (awards No.~DE-SC0019040 and No.~DE-SC0024220) and NSF QLCI (award No.~OMA-2120757). 
T.C.M. and A.V.G.~were also supported in part by AFOSR MURI, NSF STAQ program, DARPA SAVaNT ADVENT, ARL (W911NF-24-2-0107), and NQVL:QSTD:Pilot:FTL.
T.C.M., A.M.C., and A.V.G.~also acknowledge support from the U.S.~Department of Energy, Office of Science, Accelerated Research in Quantum Computing, Fundamental Algorithmic Research toward Quantum Utility (FAR-Qu).
T.C.M. and A.V.G.~also acknowledge support from the U.S.~Department of Energy, Office of Science, National Quantum Information Science Research Centers, Quantum Systems Accelerator.   
\end{acknowledgments}

\bibliographystyle{apsrev4-1-title}
\bibliography{Yuan}

\clearpage

\onecolumngrid
\makeatletter
\setcounter{MaxMatrixCols}{10}

\begin{center} 
{\large \bf Supplemental Material for: Time Independence Does Not Limit Information Flow. I. The Free-Particle Case}
\end{center}

\setcounter{secnumdepth}{2} 
\renewcommand{\thesection}{S\arabic{section}} 
\renewcommand{\thesubsection}{\Alph{subsection}}

\setcounter{equation}{0}
\renewcommand{\theequation}{S\arabic{equation}}
\setcounter{table}{0}
\renewcommand{\thetable}{S\arabic{table}}
\setcounter{figure}{0}
\renewcommand{\thefigure}{S\arabic{figure}}

In this Supplemental Material, we provide perturbation theory calculations and a rigorous upper bound on state-transfer infidelity (Section~\ref{sec:transfer_infidelity}), a complete proof of state-transfer time scaling (Section~\ref{sec:proof_scaling}), and another time-independent long-range quantum state-transfer protocol that can be implemented in current experimental platforms (Section~\ref{sec:another_protocl}).

\section{Perturbation theory calculations and rigorous upper bound on state-transfer infidelity}
\label{sec:transfer_infidelity}

\subsection{Perturbation theory calculations}
\label{subsec:perturbcalc}

In this subsection, we derive the state-transfer infidelity of our time-independent tunneling protocol for $\alpha \geq d/2$ [as shown in Fig.~1(c)--(f) of the main text] using standard perturbation theory.

By diagonalizing the hopping Hamiltonian of the state transfer channel shown in Fig.~1(e) of the main text, $\sum_{j=0}^{l-1} a^j (\ket{j} \bra{j+1} + \hc) + \sum_{j=l}^{2l-1} a^{2l-1-j} (\ket{j} \bra{j+1} + \hc)$ (recall that $a = 2^{d-\alpha}$, and we abbreviate $\ket{\text{col } j}$ as $\ket{j}$ for brevity),
we obtain the following Hamiltonian:
\begin{equation}
H = \sum_{k=0}^{2 l} E_k \ket{k}\bra{k}  + g \sum_{k=0}^{2 l}  t_k^{(0)} \left[ \ket{X} \bra{k} + (-1)^k \ket{k} \bra{Y}  + \hc \right].
\label{Eq:Fourier_Ham_1D}
\end{equation}



When the coupling strength $gt_k^{(0)}$ is small enough compared to the energy detuning $E_{k \neq \ell}$ between the resonant and off-resonant levels, up to the zeroth order in $g t_k^{(0)} / E_k$, we neglect all the off-resonant eigenmodes. For the three resonant levels $\{\ket{X}, \ket{k=l}, \ket{Y}\}$, the tunneling Hamiltonian reads $H = g t_l^{(0)} [\ket{X} \bra{k=l} + (-1)^l \ket{k=l}\bra{Y} + \hc ]$. Since
\begin{equation}
\exp\left[- i g t_l^{(0)} T
\begin{pmatrix}
0 & 1 & 0\\
1 & 0 & \pm 1\\
0 & \pm 1 & 0
\end{pmatrix}\right] \begin{pmatrix}
1 \\
0 \\
0 
\end{pmatrix} 
=
\begin{pmatrix}
[1+\cos(\sqrt{2}g t_l^{(0)} T )]/2 \\
-i \sin(\sqrt{2}g t_l^{(0)} T) / \sqrt{2}\\
\mp[1 - \cos(\sqrt{2}g t_l^{(0)} T)]/2
\end{pmatrix},
\end{equation}
after evolution time $T = \pi / (\sqrt{2}g t_l^{(0)})  = \pi / \Omega_l$ with $\Omega_k = \sqrt{2} g t_k^{(0)}$, the single particle will tunnel from site $X$ to site $Y$, and vice versa.

Next, in order to compute the first-order correction, we split the Hamiltonian Eq.~\eqref{Eq:Fourier_Ham_1D} into two parts $H = H_+ + H_-$~\cite{Yao2013Quantum}. For the convenience of later calculations, we assume that $l$ is even and that the resonant level lies in $H_+$. Analogous steps apply to odd $l$. We have
\begin{eqnarray}
H_+ &=&  \Omega_l (\ket{+} \bra{k=l} + \hc ) + \sum_{k\neq l\  \text{even}}E_k \ket{k}\bra{k} + \sum_{k\neq l\  \text{even}} \Omega_k (\ket{+} \bra{k} + \hc ),\\
H_- &=& \sum_{k\ \text{odd}}E_k \ket{k}\bra{k} + \sum_{k\ \text{odd}} \Omega_k (\ket{-} \bra{k} + \hc ),
\end{eqnarray}
where $\ket{\pm} = (\ket{X}\pm \ket{Y})/\sqrt{2}$. Note that $H_+$ and $H_-$ commute since they have no common support. 


For the $H_-$ part, which does not include the resonant level $\ket{k=l}$, first-order perturbation theory relates the unperturbed and perturbed (denoted by a tilde) eigenstates:
\begin{equation}
\ket{-} \approx \left[1 - \frac{1}{2} \sum_{k\  \text{odd}} \left(\frac{\Omega_k}{E_k}\right)^2 \right] \ket{\Tilde{-}} + \sum_{k\  \text{odd}} \frac{\Omega_k}{E_k} \ket{\Tilde{k}}.
\end{equation}

The perturbed eigenvalues of $\ket{\Tilde{-}}$ and $\ket{\Tilde{k}}$ (up to second order in $\Omega_k / E_k$) are $0$ and $E_k +\Omega_k^2 / E_k$, respectively.
Since the single-particle spectrum $\{E_k\}_{k=0}^{2l}$ is symmetric about zero and $|\Omega_k|=|\Omega_{2l-k}|$, the Stark shift of $\ket{-}$ caused by the upper half of the spectrum cancels with the Stark shift caused by the lower half of the spectrum. 
We then find that
\begin{equation}
\bra{-} e^{-iH_{-}T} \ket{-} \approx 1 - \sum_{k\ \text{odd}}   \left(\frac{\Omega_k}{E_k}\right)^2 (1 - e^{-i E_k T}).
\label{Eq:--evolution}
\end{equation}

For the $H_+$ part, since $\ket{+}$ and $\ket{k=l}$ are coupled with strength $\Omega_l$, the unperturbed eigenstates are 
$\ket{s} = \frac{1}{\sqrt{2}}\left(\ket{+} + \ket{k=l} \right)$,
with energy $\Omega_l$,
and $\ket{a} = \frac{1}{\sqrt{2}}\left(\ket{+} - \ket{k=l} \right)$, with energy $-\Omega_l$.
We rewrite $H_+$ as
\begin{equation}
H_+ =  \Omega_l \ket{s}\bra{s}-\Omega_l \ket{a}\bra{a} +
\sum_{k\neq l\  \text{even}}E_k \ket{k}\bra{k} +  \sum_{k\neq l\  \text{even}} \frac{\Omega_k}{\sqrt{2}} (\ket{s} \bra{k} +\ket{a} \bra{k} + \hc ).
\label{Eq:saH+}
\end{equation}

According to first-order perturbation theory, we find that
\begin{equation}
\ket{+} \approx \left[ 1 - \frac{1}{2}\sum_{k\neq l\ 
 \text{even}} \left(\frac{\Omega_k}{E_k}\right)^2 \right] \frac{\ket{\Tilde{s}} + \ket{\Tilde{a}}}{\sqrt{2}} + \sum_{k\neq l\ 
 \text{even}} \frac{\Omega_k }{E_k} \ket{\Tilde{k}}.
\end{equation}


The perturbed eigenvalues of $\ket{\Tilde{s}}$, $\ket{\Tilde{a}}$, and $\ket{\Tilde{k}}$ (up to second order in $\Omega_k / E_k$) are $\Omega_l ( 1 - \sum_{k=0}^{l-1} \Omega_k^2/E_k^2)$, $-\Omega_l ( 1 - \sum_{k=0}^{l-1} \Omega_k^2/E_k^2)$, and $E_k + \Omega_k^2 / E_k$, respectively (also note the Stark shift cancellation on $\ket{s}$ and $\ket{a}$).
After the evolution time $T=\pi/\Omega_l$, we find that
\begin{equation}
\bra{+} e^{-iH_{+}T} \ket{+} \approx - 1 + \sum_{k\neq l\  \text{even}}  \left(\frac{\Omega_k}{E_k}\right)^2 (1 + e^{-i E_k T}).
\label{Eq:++evolution}
\end{equation}

Combining Eqs.~\eqref{Eq:--evolution} and \eqref{Eq:++evolution}, the infidelity of the quantum state transfer can be computed as
\begin{align}
\epsilon & = 1 - |\bra{Y} e^{-iHT} \ket{X}|^2  \nonumber \\
& = 1 - \frac{1}{4}|\bra{+} e^{-iH_+ T} \ket{+} - \bra{-} e^{-iH_- T} \ket{-}|^2  \nonumber \\
& \approx 1 - \left|1 - \sum_{k \neq l}\left(\frac{\Omega_k}{E_k}\right)^2 \frac{1 + (-1)^k e^{ -i E_k T }}{2}\right|^2  \nonumber \\
& \approx \sum_{k \neq l}\left(\frac{\Omega_k}{E_k}\right)^2 \left[1 + (-1)^k \cos E_k T \right],
\end{align}
where $T = \pi/ \Omega_l$. To achieve state-transfer fidelity $1-\epsilon$, the quantum state-transfer time scales as 
\begin{equation}
T = \mathcal{O}\left[ \sqrt{ \frac{1}{\epsilon} \sum_{k \neq l} \left(\frac{\Omega_k/\Omega_l}{E_k}\right)^2 }\  \right] = \mathcal{O}\left[ \sqrt{ \frac{1}{\epsilon} \sum_{k \neq l} \left(\frac{t_k^{(0)}/t_l^{(0)}}{E_k}\right)^2 }\  \right].
\end{equation}




\subsection{The rigorous upper bound}
\label{subsec:rigorous_upper_bound}

In this subsection, we rigorously prove that the infidelity of our time-independent state-transfer protocol for $\alpha \geq d/2$ is upper bounded by $\mathcal{O}(\sum_{k\neq l}\Omega_k^2 / E_k^2)$.

In the Hamiltonian Eq.~\eqref{Eq:Fourier_Ham_1D}, the single-particle spectrum $\{E_k\}_{k=0}^{2l}$ is symmetric about zero ($E_l=0$), since by flipping the signs of all the odd-site coefficients in $\ket{k}$ [$t_{k}^{(2j)}\to t_{k}^{(2j)}$, $t_{k}^{(2j+1)}\to -t_{k}^{(2j+1)}$], we obtain another eigenstate with eigenenergy $-E_k$. 
After being weakly coupled to the two external sites $X$ and $Y$, the perturbed eigenvalues of $H$ are still symmetric about zero, since the operator $ -\ket{X}\bra{X} + \sum_{j=0}^{2l} (-1)^j \ket{j} \bra{j} - \ket{Y}\bra{Y}$ anticommutes with $H$. Therefore, the perturbed eigenvalues obey $E_0(g)>E_1(g)>\cdots>E_{l-1}>E_s(g)>E_-(g)>E_a(g)>E_{l+1}(g)>\cdots>E_{2l}(g)$, where $E_-(g)=0$, $E_s(g)= -E_a(g)$, and $E_k(g) = - E_{2l-k}(g)$ for $k=0,1,\ldots,l-1$. 

Following the notation used in the previous subsection, the state-transfer infidelity is $\epsilon = 1 - |\bra{Y} e^{-iHT} \ket{X}|^2 = 1 - |\bra{+} e^{-iH_+ T} \ket{+} - \bra{-} e^{-iH_- T} \ket{-}|^2 / 4 $. We denote
\begin{equation}
\bra{+} e^{-iH_+ T} \ket{+} = -1 + \delta_+, \quad \bra{-} e^{-iH_- T} \ket{-} = 1 + \delta_-,
\end{equation}
and thus find that 
\begin{equation}
\epsilon = 1 - \left|1 - \frac{\delta_+ - \delta_-}{2}\right|^2 = \mathrm{Re}(\delta_+ - \delta_-) - \frac{|\delta_+ - \delta_-|^2}{4} \leq |\delta_+| + |\delta_-|.
\end{equation}
The central task becomes upper bounding $|\delta_+|$ and $|\delta_-|$.

For the $H_-$ part, by symmetry of the spectrum, the perturbed eigenvalue $E_-(g)$ (including corrections at all orders) of the perturbed eigenstate $\ket{\Tilde{-}}$ is exactly zero, which motivates conjecturing that the perturbed eigenstate $\ket{\Tilde{-}}$ can be exactly written as
\begin{equation}
\ket{\Tilde{-}} = \frac{\ket{-} - \sum_{k\  \text{odd}}\frac{\Omega_k}{E_k}\ket{k}}{\sqrt{1 + \sum_{k\  \text{odd}} \Omega_k^2 / E_k^2 }}. 
\end{equation}
By using the relation $\sum_{k\  \text{odd}} \Omega_k^2 / E_k = 0$, one can verify that indeed $H_- \ket{\Tilde{-}} = 0$. Then 
$|\langle - \ket{\Tilde{-}}|^2 = 1 /(1 + \sum_{k\  \text{odd}} \Omega_k^2 / E_k^2)$.

In order to bound the $\delta_-$ term, using
\begin{equation}
\bra{-} e^{-iH_- T} \ket{-} = \langle -\ket{\Tilde{-}}\bra{\Tilde{-}}-\rangle + \sum_{k\  \text{odd}} e^{-iE_k(g)T}\langle -\ket{\Tilde{k}}\bra{\Tilde{k}}-\rangle,
\end{equation}
we find that
\begin{align}
|\delta_-| & = \left| - \frac{\sum_{k\  \text{odd}} \Omega_k^2/E_k^2 }{1 + \sum_{k\  \text{odd}} \Omega_k^2/E_k^2 } + \sum_{k\  \text{odd}} e^{-iE_k(g)T}\langle -\ket{\Tilde{k}}\bra{\Tilde{k}}-\rangle \right| \nonumber \\    
& \leq \sum_{k\  \text{odd}} \frac{\Omega_k^2}{E_k^2} +  \sum_{k\  \text{odd}} \langle -\ket{\Tilde{k}}\bra{\Tilde{k}}-\rangle \nonumber \\  
& = \sum_{k\  \text{odd}} \frac{\Omega_k^2}{E_k^2} + \langle -| (I -\ket{\Tilde{-}}\bra{\Tilde{-}})|-\rangle \nonumber \\
& \leq 2 \sum_{k\  \text{odd}} \frac{\Omega_k^2}{E_k^2} .
\label{Eq:delta-bound}
\end{align}

Similarly, in order to bound the $\delta_+$ term, according to Eq.~\eqref{Eq:saH+},
\begin{equation}
\bra{+} e^{-iH_+ T} \ket{+} = \langle +\ket{\Tilde{s}}\bra{\Tilde{s}}+\rangle e^{-iE_s(g)T} + \langle +\ket{\Tilde{a}}\bra{\Tilde{a}}+\rangle e^{-i E_a(g)T} + \sum_{k\neq l\  \text{even}}  e^{-i E_k(g) T}\langle +\ket{\Tilde{k}}\bra{\Tilde{k}}+\rangle.
\end{equation}

Using $E_s(g) = - E_a(g)$ and $|\langle +\ket{\Tilde{s}}| = |\langle +\ket{\Tilde{a}}|$, and taking $T = \pi/E_s(g)$, we find that
\begin{align}
|\delta_+| & = \left|2 \left(\frac{1}{2} - |\langle +\ket{\Tilde{s}}|^2 \right) + \sum_{k\neq l\  \text{even}}  e^{-i E_k(g)T}\langle +\ket{\Tilde{k}}\bra{\Tilde{k}}+\rangle  \right| \nonumber \\
& \leq 2 \left|\frac{1}{2} - |\langle +\ket{\Tilde{s}}|^2\right| + \bra{+} (I - \ket{\Tilde{s}}\bra{\Tilde{s}} - \ket{\Tilde{a}}\bra{\Tilde{a}}) \ket{+} \nonumber \\
& \leq 4 \left|\frac{1}{2} - |\langle +\ket{\Tilde{s}}|^2\right|.
\end{align}
The remaining task is to bound the difference between $|\langle +\ket{\Tilde{s}}|^2$ and $1/2$.

Before proceeding, we justify the necessity of the somewhat involved calculations below. Since the state-transfer time scales inversely with the coupling constant $g$, time-dependent perturbation methods such as the Dyson series---which contains secular terms---yield sub-optimal error bounds. Similarly, in the upcoming analysis, simply truncating the perturbation series and bounding the higher-order terms would produce error bounds that are significantly looser than what we find numerically and through the calculations in subsection \ref{subsec:perturbcalc}.



Using the Riesz projection \cite[p. 67]{Kato2013Perturbation}
\begin{equation}
\ket{\Tilde{s}} \bra{\Tilde{s}} = \frac{1}{2\pi i} \oint_\Gamma (z-H_+)^{-1} dz
\end{equation}
onto the eigenstate $\ket{\Tilde{s}}$, where $\Gamma$ is a contour that encloses $E_s(g)$ (near $\Omega_l$) but no other eigenvalues of $H_+$, we have
\begin{equation}
|\langle +\ket{\Tilde{s}}|^2 = \frac{1}{2\pi i} \oint_\Gamma  \bra{+}(z-H_+)^{-1}\ket{+} dz.
\end{equation}

We split $H_+$ into the diagonal part $H_0 = \Omega_l \ket{s}\bra{s}-\Omega_l \ket{a}\bra{a} +
\sum_{k\neq l\  \text{even}}E_k \ket{k}\bra{k}$ and the off-diagonal part $V = \sum_{k\neq l\  \text{even}} \frac{\Omega_k}{\sqrt{2}} (\ket{s} \bra{k} +\ket{a} \bra{k} + \hc )$. Then 
\begin{equation}
(z-H_+)^{-1} = (z-H_0)^{-1} \left[1 - V(z-H_0)^{-1} \right]^{-1}.
\end{equation}

We have
\begin{equation}
\bra{+} (z-H_0)^{-1} = \frac{1}{\sqrt{2}}\left(\frac{1}{z-\Omega_l}\bra{s} + \frac{1}{z+\Omega_l}\bra{a}\right).
\end{equation}

For the other half $[1 - V(z-H_0)^{-1}]^{-1} \ket{+} $, we observe that
\begin{equation}
\frac{1}{1 - V (z-H_0)^{-1}} = 1 + V\frac{1}{z-H_0} + V\frac{1}{z-H_0} V\frac{1}{z-H_0} + \cdots.
\end{equation}
We only consider the terms containing even numbers of $V$s, and find that
\begin{equation}
\left(V\frac{1}{z-H_0} \right)^2 \ket{+} = \frac{z }{z^2 - \Omega_l^2} \sum_{k\neq l\  \text{even}} \frac{ \Omega_k^2}{z-E_k} \ket{+}.
\end{equation}

Putting everything together, we have
\begin{equation}
\bra{+}(z-H_+)^{-1}\ket{+}  = \frac{ \frac{z}{z^2 - \Omega_l^2} }{ 1- \frac{z}{z^2 - \Omega_l^2} \sum_{k\neq l\  \text{even}}\frac{\Omega_k^2}{z-E_k} } =  \frac{ \frac{z}{z^2 - \Omega_l^2} }{ 1 + \frac{z^2}{z^2 - \Omega_l^2} \sum_{k\neq l\  \text{even}}\frac{\Omega_k^2}{E_k^2 - z^2} },
\end{equation}
where in the last equality we used the symmetry of the spectrum about zero. 

Since
\begin{equation}
\frac{1}{2\pi i}\oint_\Gamma \frac{z}{z^2-\Omega_l^2} dz = \frac{1}{2},
\end{equation}
we have
\begin{equation}
|\langle +\ket{\Tilde{s}}|^2 = \frac{1}{2} - \frac{1}{2\pi i} \oint_\Gamma f(z)dz
\end{equation}
where
\begin{equation}
f(z) = \frac{z^3}{(z^2 - \Omega_l^2)^2} \frac{ \sum_{k\neq l\  \text{even}}\frac{\Omega_k^2}{E_k^2 - z^2} }{ 1 + \frac{z^2}{z^2 - \Omega_l^2} \sum_{k\neq l\  \text{even}}\frac{\Omega_k^2}{E_k^2 - z^2} }.
\end{equation}

Next we choose an appropriate contour $\Gamma$ to upper bound the absolute value of $f(z)$. According to the proof of Theorem 1 in Ref.~\cite{Mathias1998Quadratic}, the difference between the perturbed eigenvalue $E_s(g)$ and the unperturbed eigenvalue $\Omega_l$ is upper bounded by
\begin{equation}
|E_s(g) - \Omega_l| \leq \Omega_l \sum_{k\neq l\  \text{even}} \frac{\Omega_k^2}{E_k^2 - \Omega_l^2}.
\label{Eq:distance_Esg_Omegal}
\end{equation}
Furthermore, according to Cauchy's eigenvalue interlacing theorem, $E_l=0 \leq E_s(g) \leq E_{l-2} \leq E_{l-2}(g)$.
As long as $E_{l-2}\ge 2 \Omega_l$ and $\sum_{k\neq l\  \text{even}} \Omega_k^2 / E_k^2 < 3/4$, the contour $\Gamma$ can be chosen as $z=\Omega_l(1+e^{i\theta})$, $\theta\in[0,2\pi)$ to only enclose $E_s(g)$ but no other eigenvalues of $H_+$. On this particular contour, we have
\begin{equation}
|f(z)| = |f(\Omega_l(1+e^{i\theta}))| = \frac{1}{\Omega_l} \frac{|1+e^{i\theta}|^3}{|2 + e^{i\theta}|^2} \times 
\frac{ \left|\sum_{k\neq l\  \text{even}}\frac{\Omega_k^2}{E_k^2 - \Omega_l^2(1+e^{i\theta})^2}\right| }{ \left| 1 + \frac{(1+e^{i\theta})^2}{e^{i\theta}(2+e^{i\theta}) } \sum_{k\neq l\  \text{even}}\frac{\Omega_k^2}{E_k^2 - \Omega_l^2(1+e^{i\theta})^2} \right|}.   
\label{Eq:f_theta}
\end{equation}


For the numerator of the term after the multiplication sign in Eq.~\eqref{Eq:f_theta},
\begin{equation}
\left| \sum_{k\neq l\  \text{even}}\frac{\Omega_k^2}{E_k^2 - \Omega_l^2(1+e^{i\theta})^2}\right| \leq \sum_{k\neq l\  \text{even}}\frac{\Omega_k^2}{E_k^2 - 4\Omega_l^2},
\end{equation}
under the condition that $E_{l-2} > 2 \Omega_l$.

For the denominator of the term after the multiplication sign in Eq.~\eqref{Eq:f_theta},
\begin{equation}
\left| 1 + \frac{(1+e^{i\theta})^2}{e^{i\theta}(2+e^{i\theta}) } \sum_{k\neq l\  \text{even}}\frac{\Omega_k^2}{E_k^2 - \Omega_l^2(1+e^{i\theta})^2} \right| \ge 1 + \sum_{k\neq l\  \text{even}} \Omega_k^2 \text{ Re}\left\{\frac{(1+e^{i\theta})^2}{e^{i\theta}(2+e^{i\theta})[E_k^2 -\Omega_l^2(1+e^{i\theta})^2]}\right\}.
\label{Eq:denominator_f(z)}
\end{equation}

We want to prove that the right-hand side of 
the inequality Eq.~\eqref{Eq:denominator_f(z)} is larger than $1$.
We denote $u = (1+e^{i\theta})^2$  and $v = e^{i\theta}(2+e^{i\theta})[E_k^2 -\Omega_l^2(1+e^{i\theta})^2]$. Then $\Re(u/v) = \Re(uv^*) / |v|^2 $ and
\begin{equation}
\Re(uv^*) = 2(1+\cos\theta) \left\{ (2+\cos\theta)[E_k^2 - \Omega_l^2 (1+2\cos\theta+\cos2\theta)] + 2\Omega_l^2 \sin^2\theta(1+\cos\theta)\right\}.
\end{equation}
As long as $E_{l-2} \ge 2 \Omega_l$, we have $\Re(uv^*) \ge 0$ for all $\theta \in [0,2\pi)$, and thus the denominator of the term after the multiplication sign in Eq.~\eqref{Eq:f_theta} is larger than $1$. 


Finally, under the condition $E_{l-2} > 2 \Omega_l$, we find that
\begin{equation}
|f(z)| \leq \frac{1}{\Omega_l} \frac{|1+e^{i\theta}|^3}{|2 + e^{i\theta}|^2}
\sum_{k\neq l\  \text{even}}\frac{\Omega_k^2}{E_k^2 - 4\Omega_l^2},      
\end{equation}
which gives
\begin{equation}
|\delta_+| 
\leq 4 \left|\frac{1}{2\pi i} \oint_\Gamma f(z)dz  \right| 
\leq 4 \sum_{k\neq l\  \text{even}}\frac{\Omega_k^2}{E_k^2 - 4\Omega_l^2} \times 0.505 \leq  3 \sum_{k\neq l\  \text{even}}\frac{\Omega_k^2}{E_k^2 }.
\label{Eq:delta+bound}
\end{equation}
In the last inequality, we further require $E_k^2 - 4\Omega_l^2 \ge 3 E_k^2 / 4$, or equivalently, $E_{l-2} \ge 4 \Omega_l$.

Combining Eq.~\eqref{Eq:delta-bound} and Eq.~\eqref{Eq:delta+bound}, we derive that the state-transfer infidelity is upper bounded by
\begin{equation}
\epsilon = 1 - |\bra{Y} e^{-iHT} \ket{X}|^2 \leq |\delta_-| + |\delta_+| \leq 3 \sum_{k\neq l} \frac{\Omega_k^2}{E_k^2} = \mathcal{O} \left(\sum_{k\neq l} \frac{\Omega_k^2}{E_k^2}\right),
\end{equation}
provided that $E_{l-2} \ge 4 \Omega_l$ and $\sum_{k\neq l} \Omega_k^2 / E_k^2 < 3/4$.
The second condition can be consistently satisfied by choosing small enough constant $\epsilon$.
We justify that the first condition can be achieved by choosing appropriate $g$ at the end of Sec.~\ref{sec:proof_scaling}. Combining these two conditions and Eq.~\eqref{Eq:distance_Esg_Omegal}, the state-transfer time is $T = \pi / E_s(g) = \mathcal{O}(1/\Omega_l) = \mathcal{O}(1/g t_l^{(0)}) $.


\section{Rigorous proof for the state-transfer time scaling}
\label{sec:proof_scaling}

\begin{figure}
\hspace*{0\textwidth}
\includegraphics[width=0.9\linewidth]{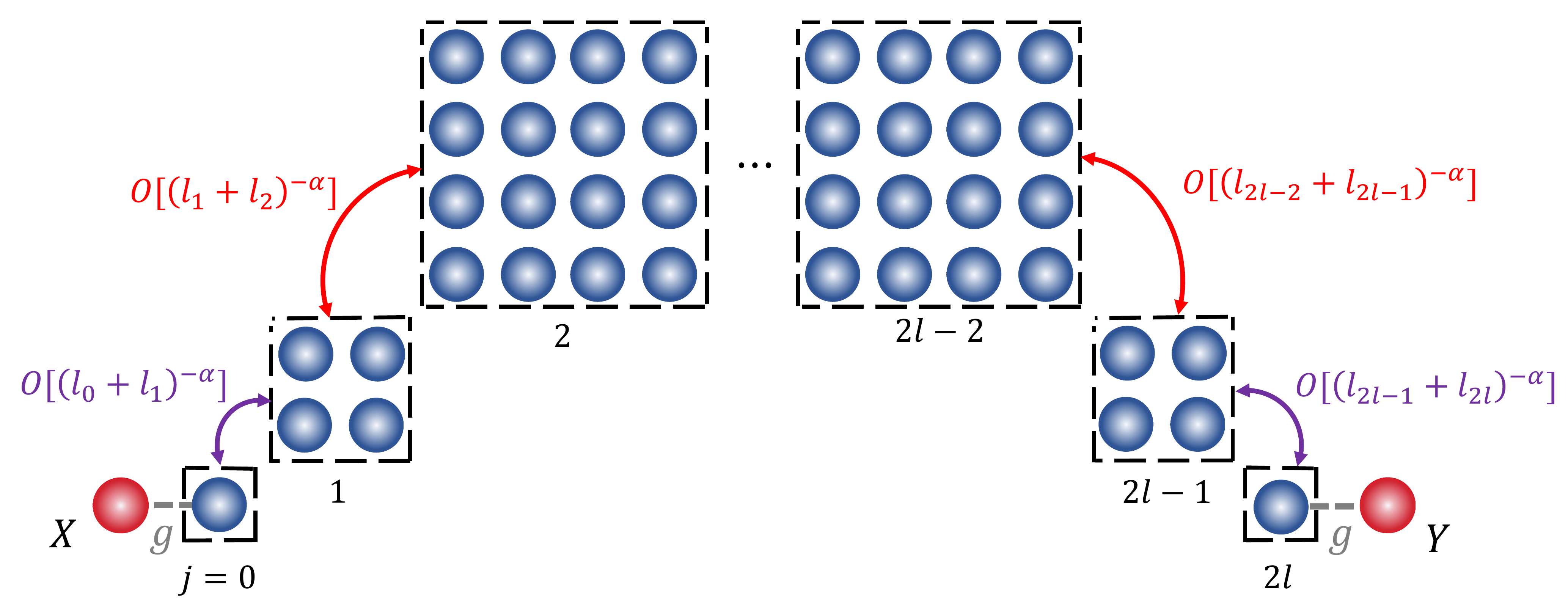} 
\caption{An illustration of the block arrangement in our time-independent quantum state-transfer protocol for $\alpha\ge d/2$. The neighboring blocks are disjoint and touch only at their corners. All the hopping strengths between sites in the $j$th block and those in the $(j+1)$st block are uniformly equal to $\bigl[ \sqrt{ d \times (l_j + l_{j+1})^2 } \bigr] ^{-\alpha}$. Here we show the $d=2$ case, and higher-dimensional cases are analogous.}
\label{fig:Illustration_d=2}
\end{figure}



In this section, we provide a complete proof of the state-transfer time scaling of our time-independent protocol for $\alpha\ge d/2$. 
Fig.~\ref{fig:Illustration_d=2} illustrates the block arrangement in our state-transfer protocol for $d=2$. We note that alternative arrangements of the disjoint blocks, such as a diagonal pattern from the bottom left to the top right or stacking all the blocks at the bottom, could also accomplish the state transfer task with the same time scaling.

Given the hopping Hamiltonian shown in Fig.~1(e) of the main text (with $a = 2^{d-\alpha}$, and abbreviating $\ket{\text{col } j}$ as $\ket{j}$ for simplicity),
\begin{equation}
H = \sum_{j=0}^{l-1} a^j (\ket{j} \bra{j+1} + \hc) + \sum_{j=l}^{2l-1} a^{2l-1-j} (\ket{j} \bra{j+1} + \hc).
\end{equation}
Due to spatial inversion symmetry, the eigenstate $\ket{k}$ with eigenenergy $E_k$ has the form 
\begin{equation}
\ket{k} = \sum_{j=0}^{l-1} t_k^{(j)} \ket{j} + t_k^{(l)} \ket{l} + (-1)^k \sum_{j=0}^{l-1} t_k^{(j)} \ket{2l - j}.
\end{equation}
Here, $(-1)^k$ corresponds to the parity of the eigenstate. We consider the case where $l$ is even (the proof for the odd case is essentially the same).




As illustrated in Sec.~\ref{sec:transfer_infidelity}, the state-transfer time of our time-independent protocol is upper bounded by
\begin{equation}
T = \mathcal{O}\left(\sqrt{\frac{1}{\epsilon} \sum_{k\neq l} \left[\frac{t_k^{(0)}/t_l^{(0)}}{E_k}\right]^2} \right).
\end{equation}
Our ultimate task is to rigorously prove that the quantity $Q = \sqrt{\sum_{k \neq l} \left(\frac{t_k^{(0)}/t_l^{(0)}}{E_k}\right)^2}$ scales as $\mathcal{O}(1)$ when $a>1$ and as $\mathcal{O}(a^{-l})$ when $a<1$. Our proof has two main parts:
(1) We employ the methods of proof by contradiction and induction to show that $E_{l-1}$ (the smallest energy gap) is lower bounded by $\Omega(1)$ when $a>1$ and $\Omega(a^{l})$ when $a<1$;
(2) By exploiting the orthonormal property of the coefficients $\{t_k^{(j)}\}$, we use several inequalities to upper bound the quantity $Q$.



We start by solving the eigenstate equation $H \ket{k} = E_k \ket{k}$, which gives us the following recurrence relations:
\begin{equation}
\begin{aligned}
E_k t_k^{(0)} &= a^0 t_k^{(1)}, \\
E_k t_k^{(1)} &= a^0 t_k^{(0)} + a^1 t_k^{(2)}, \\
E_k t_k^{(2)} &= a^1 t_k^{(1)} + a^2 t_k^{(3)}, \\
E_k t_k^{(3)} &= a^2 t_k^{(2)} + a^3 t_k^{(4)}, \\
&\hspace{0.5em}\vdots \\
E_k t_k^{(l-2)} &= a^{l-3} t_k^{(l-3)} + a^{l-2} t_k^{(l-1)}, \\
E_k t_k^{(l-1)} &= a^{l-2} t_k^{(l-2)} + a^{l-1} t_k^{(l)}, \\
E_k t_k^{(l)} &= a^{l-1} t_k^{(l-1)} + (-1)^k a^{l-1} t_k^{(l-1)}.
\label{Eq:eigenstate_recurrence_relation}
\end{aligned}
\end{equation}
 
In particular, we find that the $l$th eigenstate with $E_l = 0$ can be exactly solved, which gives
\begin{equation}
t_{l}^{(2j)} = \frac{1}{(-a)^j} \times \frac{a^{l/2}}{\sqrt{1+2(a^l-1)/(1-a^{-2})}}, \quad t_{l}^{(2j+1)} = 0.    
\end{equation}


For other eigenstates with $E_k>0$, in order to find a pattern to acquire some physical intuition, we explicitly calculate the first several coefficients:
\begin{equation}
\begin{aligned}
t_k^{(0)} &= 1, \\
t_k^{(1)} &= E_k, \\
t_k^{(2)} &= \frac{1}{-a}(1- E_k^2), \\
t_k^{(3)} &= E_k \frac{(a^2 + 1) - E_k^2 }{(-a)^3}, \\
t_k^{(4)} &= \frac{a^4 (1- E_k^2) - (a^2 + 1 ) E_k^2 + E_k^4}{(-a)^6}, \\
t_k^{(5)} &= E_k \frac{(a^8 + a^6 + a^4)  - ( a^6 + a^4 + a^2 + 1) E_k^2 + E_k^4}{a^{10}}, \\
&\hspace{0.5em}\vdots
\end{aligned}
\label{Eq:first_several_coefficients}
\end{equation}
(without loss of generality, we can take $t_k^{(0)} = 1$ and carry out a normalization after calculating all the coefficients).

\subsection{The smallest gap scales as \texorpdfstring{$\Omega(1)$}{Ω(1)} when \texorpdfstring{$a > 1$}{a>1}}

In this subsection, we prove that the smallest energy gap scales as $\Omega(1)$ when $a>1$.
Assume that there exists an eigenstate with eigenvalue $E_{k \neq \ell} =o(1)$ (i.e., vanishing as system size increases). In this subsection, we omit the subscript $k$ of $E_k$ for simplicity. Then, $E$ can be used as a small parameter to upper and lower bound the coefficients $\{t_k^{(j)}\}$ (in this subsection we write $\{t_k^{(j)}\}$ as $\{t_j\}$ for notational simplicity).


We observe from Eq.~\eqref{Eq:first_several_coefficients} that, in the large $a$ and small $E$ limit, the leading order of the coefficients approximately becomes
\begin{equation}
t_{2j} \approx (1 - E^2) / (-a)^j, \quad t_{2j+1} \approx E / (-a)^j.
\end{equation}
Based on this observation, we can use induction to see that
\begin{equation}
\frac{1 - \alpha_j E^2}{a^j} \leq |t_{2j}| \leq \frac{1 - \beta_j E^2}{a^j}, \qquad \gamma_j \frac{E}{a^j} \leq |t_{2j+1}| \leq \eta_j \frac{E}{a^j},
\label{Eq:c_j_induction_age1}
\end{equation}
where we have introduced the undetermined $E$-independent constants 
\begin{equation}
1 \leq \alpha_1 \leq \alpha_2 \leq \alpha_3 \leq \cdots, \qquad 
1 \ge \beta_1 \ge \beta_2 \ge \beta_3 \ge \cdots >0,
\end{equation}
\begin{equation}
1 \ge \gamma_1 \ge \gamma_2 \ge \gamma_3 \ge \cdots >0, \qquad 
2 \leq \eta_1 \leq \eta_2 \leq \eta_3 \leq \cdots.
\end{equation}
In addition, the signs of $\{t_j\}$ inductively obey the following pattern: 
$t_j > 0$ when $j \equiv 0,1 \text{ mod } 4$ and $t_j < 0$ when $j \equiv 2,3 \text{ mod } 4$. 



Notice that, according to Eq.~\eqref{Eq:first_several_coefficients}, $t_2, t_3$ already satisfy Eq.~\eqref{Eq:c_j_induction_age1} and the sign pattern. The recurrence relations Eq.~\eqref{Eq:eigenstate_recurrence_relation} require that 
\begin{equation}
\alpha_j +\frac{\eta_j}{a^{2j}} \leq \alpha_{j+1}, 
\quad  \beta_j +\frac{\gamma_j}{a^{2j}} \ge \beta_{j+1}, 
\quad \gamma_j + \frac{1-\alpha_{j+1} E^2}{a^{2j+2}} \ge \gamma_{j+1}, 
\quad \eta_j + \frac{1-\beta_{j+1} E^2}{a^{2j+2}} \leq \eta_{j+1}.
\label{Eq:coeff_relation_age1}
\end{equation}

We particularly take $\beta_j \equiv 1$, $\gamma_j \equiv 1$, such that the second and third inequalities in Eq.~\eqref{Eq:coeff_relation_age1} are satisfied if $1-\alpha_{l/2} E^2 \ge 0$.
Next, in order to satisfy the fourth inequality in Eq.~\eqref{Eq:coeff_relation_age1}, we take 
\begin{equation}
\eta_{j+1} = \eta_j + \frac{1}{a^{2j+2}},
\end{equation}
which gives
\begin{equation}
2 \leq \eta_1 \leq \eta_2 \leq \eta_3 \leq \cdots < 2 + \sum_{j = 1}^\infty \frac{1}{a^{2j+2}} = 2 + \frac{1}{a^2(a^2-1)}.
\end{equation}
Finally, to satisfy the first inequality in Eq.~\eqref{Eq:coeff_relation_age1}, we take 
\begin{equation}
\alpha_{j+1} = \alpha_j + \frac{2 + 1 / a^2(a^2-1)}{a^{2j}},   
\end{equation}
which similarly leads to
\begin{equation}
1 \leq \alpha_1 \leq \alpha_2 \leq \alpha_3 \leq \cdots < 1 + \left[2 + \frac{1}{a^2(a^2-1)}\right] \times \sum_{j = 1}^\infty \frac{1}{a^{2j}} = 1 + \frac{2}{a^2-1} + \frac{1}{a^2 (a^2-1)^2}.
\end{equation}
Therefore, $1-\alpha_{l/2} E^2 \ge 0$ is consistently satisfied since $E=o(1)$.

With Eq.~\eqref{Eq:c_j_induction_age1} in hand, we conclude the proof by observing the contradiction with the last recurrence relation in Eq.~\eqref{Eq:eigenstate_recurrence_relation}: if this eigenstate with the eigenvalue $E$ has odd parity, then $t_l  = 1 - \text{const}\times E^2 = 0$, contradicting the assertion that $E=o(1)$; if this eigenstate has even parity, then $E t_l = 2 a^{l-1} t_{l-1}$, implying that $t_{l-1} $ and $t_l$ have the same sign, contradicting the fact that $t_j > 0$ when $j \equiv 0,1 \text{ mod } 4$ and $t_j < 0$ when $j \equiv 2,3 \text{ mod } 4$.

\subsection{The smallest gap scales as \texorpdfstring{$\Omega(a^l)$}{Ω(aᶩ)} when \texorpdfstring{$a < 1$}{a<1}}

In this subsection, we prove that the smallest energy gap scales as $\Omega(a^l)$ when $a<1$.
We similarly begin by assuming that there exists an eigenstate with eigenvalue $E_{k \neq \ell} = o(a^l)$, and eventually arrive at a contradiction. In this subsection, we again omit the subscript $k$ of $E_k$ for simplicity.

From Eq.~\eqref{Eq:first_several_coefficients}, we observe that, in the small $a$ and small $E$ limit, the leading order of the coefficients approximately becomes (in this subsection, we also write $\{t_k^{(j)}\}$ as $\{t_j\}$ for brevity)
\begin{equation}
t_{2j} \approx \frac{1}{(-a)^j} - \frac{E^2}{(-a)^{5j-4}},
\quad t_{2j+1} \approx \frac{E}{(-a)^{3j}}.
\end{equation}


We again apply induction to see that
\begin{equation}
\frac{1}{a^j} - \alpha_j \frac{E^2}{a^{5j-4}} \leq |t_{2j}| \leq \frac{1}{a^j} - \beta_j \frac{E^2}{a^{5j-4}}, \qquad \gamma_j \frac{E}{a^{3j}} \leq |t_{2j+1}| \leq \eta_j \frac{E}{a^{3j}},
\label{Eq:c_j_induction_aless1}
\end{equation}
\begin{equation}
1 \leq \alpha_1 \leq \alpha_2 \leq \alpha_3 \leq \cdots, \qquad 
1 \ge \beta_1 \ge \beta_2 \ge \beta_3 \ge \cdots >0,
\end{equation}
\begin{equation}
1 \ge \gamma_1 \ge \gamma_2 \ge \gamma_3 \ge \cdots >0, \qquad 
2 \leq \eta_1 \leq \eta_2 \leq \eta_3 \leq \cdots.
\end{equation}
In addition, the signs of $\{t_j\}$ inductively obey the following pattern: $t_j > 0$ when $j \equiv 0,1 \text{ mod } 4$ and $t_j < 0$ when $j \equiv 2,3 \text{ mod } 4$.

Notice that, according to Eq.~\eqref{Eq:first_several_coefficients}, $t_2, t_3$ already satisfy  Eq.~\eqref{Eq:c_j_induction_aless1} and the sign pattern. The recurrence relations Eq.~\eqref{Eq:eigenstate_recurrence_relation} require that 
\begin{equation}
a^4 \alpha_j + \eta_j \leq \alpha_{j+1}, 
\quad  a^4 \beta_j + \gamma_j \ge \beta_{j+1}, 
\quad 1 + a^2 \gamma_j - \frac{E^2}{a^{4j}} \alpha_{j+1} \ge \gamma_{j+1}, 
\quad 1 + a^2 \eta_j - \frac{E^2}{a^{4j}} \beta_{j+1} \leq \eta_{j+1}.
\label{Eq:coeff_relation_aless1}
\end{equation}

We particularly take $\beta_j \equiv 1$, $\gamma_j \equiv 1$, such that the second and third inequalities in Eq.~\eqref{Eq:coeff_relation_aless1} are satisfied if $\alpha_j \leq a^{4j-2} / E^2 $. Next, in order to satisfy the forth inequality in Eq.~\eqref{Eq:coeff_relation_aless1}, we take 
\begin{equation}
\eta_{j+1} = a^2 \eta_j + 1.
\end{equation}
If $a^2 \leq 1/2$, we take $\eta_j \equiv 2$; If  $a^2 > 1/2$, we obtain
\begin{equation}
\eta_j = - a^{2(j-1)}\left(\frac{1}{1-a^2} - 2\right) + \frac{1}{1-a^2}.
\end{equation}
In both cases, we obtain
\begin{equation}
2 \leq \eta_1 \leq \eta_2 \leq \eta_3 \leq \cdots \leq \eta^* =\max\left\{ 2, \frac{1}{1- a^2} \right\}.
\end{equation}

Finally, to satisfy the first inequality in Eq.~\eqref{Eq:coeff_relation_aless1}, we take 
\begin{equation}
\alpha_{j+1} = a^4 \alpha_j + \eta^*,  
\end{equation}
which gives
\begin{equation}
\alpha_j = - a^{4(j-1)}\left[\frac{\eta^*}{1-a^4} - 1\right] + \frac{\eta^*}{1-a^4}. 
\end{equation}
We thus find that
\begin{equation}
1 \leq \alpha_1 \leq \alpha_2 \leq \alpha_3 \leq \cdots < \frac{\eta^*}{1-a^4},
\end{equation}
which consistently satisfy $\alpha_j \leq a^{4j-2} / E^2 $ since $j \leq l/2$ and $E=o(a^l)$.

With Eq.~\eqref{Eq:c_j_induction_aless1} in hand, we conclude the proof by observing the contradiction with the last recurrence relation in Eq.~\eqref{Eq:eigenstate_recurrence_relation}: if this eigenstate with eigenvalue $E$ has odd parity, then $t_l = 1 - \text{const}\times E^2 / a^{2l - 4}= 0$,
contradicting the assertion that $E = o(a^l)$; if this eigenstate has even parity, then $E t_l = 2 a^{l-1} t_{l-1}$, implying that $t_{l-1} $ and $t_l$ have the same sign, contradicting the fact that $t_j > 0$ when $j \equiv 0,1 \text{ mod } 4$ and $t_j < 0$ when $j \equiv 2,3 \text{ mod } 4$.

\subsection{Exploiting the orthonormality to upper bound the quantity \texorpdfstring{$Q$}{Q}}

For the $E_l=0$ eigenstate, after normalization, we have
\begin{equation}
1 / t_l^{(0)} = \sqrt{a^{-l} + 2(1-a^{-l}) / ( 1 - a^{-2})}.
\end{equation}

For the $a>1$ case, $1/t_l^{(0)} = \mathcal{O}(1)$ and  $E_k = \Omega(1)$, so
\begin{equation}
\sum_{k\neq l} \left[\frac{t_k^{(0)}/t_l^{(0)}}{E_k}\right]^2 \leq \text{const} \times \sum_{k\neq l} \left[ t_k^{(0)} \right]^2 = \mathcal{O}(1).
\end{equation}

For the $a<1$ case, since $1/t_l^{(0)} = \mathcal{O}(a^{-l/2})$, naively replacing all the $\{E_k\}$ by $\Omega(a^l)$ will not give the desired bound. According to the recurrence relations of the eigenstates Eq.~\eqref{Eq:eigenstate_recurrence_relation}, we observe that
\begin{align}
t_k^{(0)} & = E_k t_k^{(1)} - a t_k^{(2)} = E_k \left[ t_k^{(1)} - \frac{t_k^{(3)}}{a} \right] + a^2 t_k^{(4)} = E_k \left[ t_k^{(1)} - \frac{t_k^{(3)}}{a} + \frac{t_k^{(5)}}{a^2} \right] - a^3 t_k^{(6)} \nonumber \\
& = E_k \left[ t_k^{(1)} - \frac{t_k^{(3)}}{a} + \frac{t_k^{(5)}}{a^2} + \cdots + \frac{t_k^{(l-1)}}{(-a)^{l/2-1}} \right] + (-a)^{l/2} t_k^{(l)} . 
\label{Eq:an_equality_for_t0}
\end{align}

Then we derive that
\begin{align}
\left(\frac{t_k^{(0)}}{E_k}\right)^2  & \leq 2 \left[ \left( t_k^{(1)} - \frac{t_k^{(3)}}{a} + \frac{t_k^{(5)}}{a^2} + \cdots + \frac{t_k^{(l-1)}}{(-a)^{l/2-1}} \right)^2 + a^l \frac{(t_k^{(l)})^2}{E_k^2} \right] \nonumber \\
& \leq 2 \left[ \left( t_k^{(1)} - \frac{t_k^{(3)}}{a} + \frac{t_k^{(5)}}{a^2} + \cdots + \frac{t_k^{(l-1)}}{(-a)^{l/2-1}} \right)^2 + \text{const} \times a^{-l} (t_k^{(l)})^2 \right],    
\end{align}
where we use $E_k = \Omega(a^l)$.

Finally,
\begin{align}
\sum_{k\neq l} \left[\frac{t_k^{(0)}/t_l^{(0)}}{E_k}\right]^2 & \leq 2 a^{-l} \sum_{k\neq l}\left[ t_k^{(1)} - \frac{t_k^{(3)}}{a} + \frac{t_k^{(5)}}{a^2} + \cdots + \frac{t_{k}^{(l-1)}}{(-a)^{l/2-1}}  \right]^2 + \text{const} \times a^{-2l}
 \nonumber \\
& = 2 a^{-l} \sum_{k\neq l} \left[ (t_k^{(1)})^2 + \frac{(t_k^{(3)})^2}{a^2} + \frac{(t_k^{(5)})^2}{a^4} + \cdots + \frac{(t_k^{(l-1)})^2}{a^{l-2}}  \right] + \text{const} \times a^{-2l} \nonumber \\
& = 2 a^{-l} \times \left(1 + \frac{1}{a^2} + \frac{1}{a^4} + \cdots + \frac{1}{a^{l-2}}\right) + \text{const} \times a^{-2l} \nonumber \\
& = \mathcal{O}(a^{-2l}),
\end{align}
where in the second step, we have utilized the orthonormality relations:  $\langle p | q \rangle = \sum_{k=0}^{2l} t_k^{(p)} t_k^{(q)} = \sum_{k\neq l} t_k^{(p)} t_k^{(q)} = \delta_{pq} $ for $p, q$ odd [note that $t_{l}^{(2 j+1)} = 0$].



As promised in Sec.~\ref{sec:transfer_infidelity}, we justify that the condition $E_{l-2} \ge 4 \Omega_l$ can be consistently satisfied with the scaling behaviors of the state-transfer times:
\begin{enumerate}[(1)]
\item When $a > 1$, we have shown that $E_{l-2} \ge C_1$, $t_l^{(0)} = C_2$. As long as $g \leq C_1/ (4\sqrt{2}C_2)$, $E_{l-2} \ge 4 \Omega_l$, and this choice of $g$ is consistent with the fact that $T=\pi/(\sqrt{2}g t_l^{(0)}) = \mathcal{O}(1)$.
\item When $a = 1$, $E_{l-2} = C_3 / l $, $ t_l^{(0)} = C_4 / \sqrt{l}$. As long as $g \leq C_3/ (4 \sqrt{2}C_4 \sqrt{l})$, $E_{l-2} \ge 4 \Omega_l$, and this choice of $g$ is consistent with the fact that $T=\pi/(\sqrt{2}g t_l^{(0)}) =\mathcal{O}(l)$.
\item  When $a < 1$, $E_{l-2} \ge C_5 a^l $, $ t_l^{(0)} = C_6 a^{l/2}$. As long as $g \leq [ C_5/ (4\sqrt{2}C_6) ] a^{l/2}$, $E_{l-2} \ge 4 \Omega_l$, and this choice of $g$ is consistent with the fact that $T=\pi/(\sqrt{2}g t_l^{(0)})=\mathcal{O}(a^{-l})$.
\end{enumerate}
Here $C_1,C_2,C_3,C_4,C_5,C_6$ are $l$-independent constants.

\section{Another time-independent long-range quantum state-transfer protocol}
\label{sec:another_protocl}

The two optimal time-independent state-transfer protocols presented in the main text require that the hopping strength between each two neighboring blocks to be uniform, which poses a challenge for experimental realization.
In this section, we present another time-independent free-particle state-transfer protocol with hopping strengths exactly following $J_{ij} = J_0 / r_{ij}^\alpha$, which may be easier to implement in current atomic and molecular platforms. Although suboptimal, this protocol still provides a speed-up over short-range protocols. 


\subsection{The \texorpdfstring{$d=1$}{d=1} case}

\begin{figure}
\hspace*{0\textwidth}
\includegraphics[width=1\linewidth]{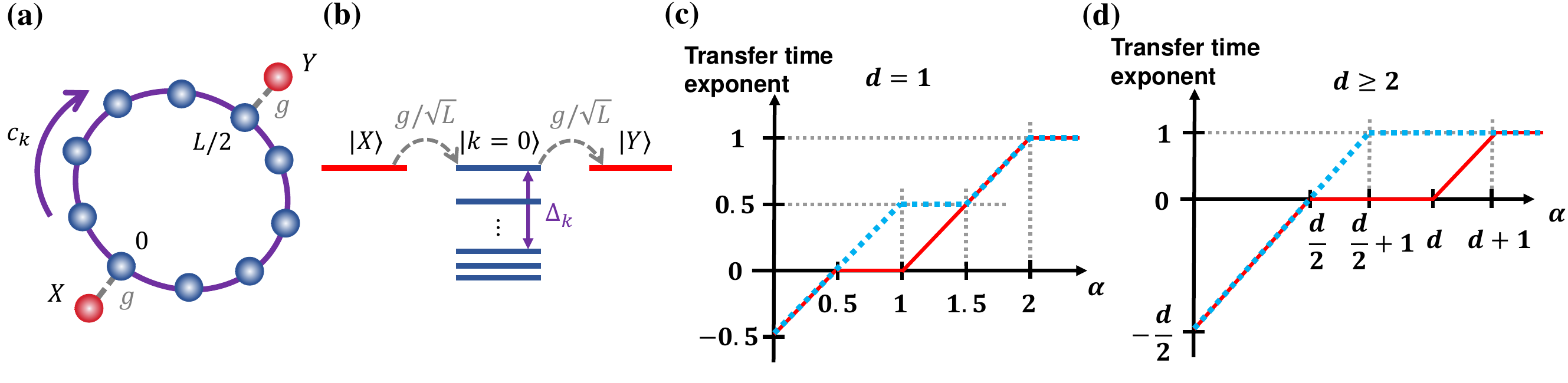} 
\caption{Summary of the time-independent free-particle state-transfer protocol with $J_{ij}$ = $J_0/r_{ij}^\alpha$. (a) An illustration showing a single particle hopping on a one-dimensional ring and the two external sites coupled to the transfer channel. (b) The single particle tunnels from  state $\ket{X}$ to state $\ket{Y}$ through momentum eigenmode $\ket{k=0}$. Coupling to other off-resonant momentum eigenmodes will result in nonzero transfer infidelity. (c,d) The power-law scaling exponent of the state-transfer time $T$ with respect to the distance $L$, for different $\alpha$ in $d=1$ [(c)] and $d\ge 2$ [(d)]. The red solid line denotes the optimal time-dependent free-particle state-transfer protocol (saturating the free-particle Lieb-Robinson bounds). The cyan dotted line denotes the results of our time-independent tunneling protocol with $J_{ij}$ = $J_0/r_{ij}^\alpha$. }
\label{fig:Fig_S1}
\end{figure}

In this subsection, we consider the $d=1$ case, which we extend to higher spatial dimensions in the next subsection. As shown in Fig.~\ref{fig:Fig_S1}(a), a one-dimensional ring lattice with $L$ sites serves as the quantum state-transfer channel. 
We attach two external sites $X$ and $Y$ to the diametrically opposite sites $0$ and $L/2$ (we assume henceforth that $L$ is even). We take the following time-independent free-particle Hamiltonian with long-range hopping to transfer the population from site $X$ to site $Y$:
\begin{equation}
H = \sum_{i \neq j}  J_{i j} c^\dagger_i c_j  + g \left(c_X^\dagger c_0 + c_{L/2}^\dagger c_{Y} + \hc \right) + \mu \left(c_X^\dagger c_X + c_{Y}^\dagger c_{Y} \right).
\label{Eq:hopping_Hamiltonian}
\end{equation}
Here $J_{ij} = J_0 / r_{ij}^\alpha $, where $r_{ij} = \min\{|i-j|, L - |i-j|\}$ and where we set $J_0=1$. The parameter $\mu$ is the chemical potential of sites $X$ and $Y$. 




Quantum state transfer is achieved by a quantum tunneling process similar to that in the main text: we tune the energy of the two external sites to be resonant with the $\ket{k = 0}$ momentum eigenmode of the transfer channel, such that the single particle tunnels from site $X$ to site $Y$ (and vice versa) mediated by state $\ket{k = 0}$, as illustrated in Fig.~\ref{fig:Fig_S1}(b). When the coupling strength $g$ is small enough compared to the energy differences between $\ket{k = 0}$ and other momentum eigenmodes, the transfer infidelity caused by the off-resonant levels can be bounded above by a desired constant $\epsilon$. In the following, we numerically and analytically evaluate the scaling behavior of the state-transfer time $T$ with respect to different $L$ and $\alpha$.

The translation-invariant hopping Hamiltonian describing the transfer channel in  Eq.~\eqref{Eq:hopping_Hamiltonian} can be diagonalized by a Fourier transformation: $c_k = \frac{1}{\sqrt{L}}\sum_{j = 0}^{L-1} e^{ i 2 \pi j k / L } c_j$, $k = 0, 1, \ldots, L-1$. The resulting single-particle spectrum of the transfer channel is 
\begin{equation}
E_k = 2 \sum_{j=1}^{L/2 - 1} \frac{\cos( 2 \pi k j / L)}{j ^ \alpha}  + \frac{(-1)^ {k}}{(L/2) ^ \alpha}.   
\end{equation}
We shift the energy of the $\ket{k=0}$ momentum eigenmode to be zero and rewrite the Hamiltonian Eq.~\eqref{Eq:hopping_Hamiltonian} as
\begin{equation}
H = \sum_{k=0}^{L-1} \Delta_k c_k^\dagger c_k  + \frac{g}{\sqrt{L}} \sum_{k=0}^{L-1}  \left[ c_X^\dagger c_k + (-1)^k c_{k}^\dagger c_{Y} + \hc \right]  + \mu (c_X^\dagger c_X + c_{Y}^\dagger c_{Y}),      
\label{Eq:Fourier_Ham_d=1}
\end{equation}
where $\Delta_k = E_{k=0} - E_{k}$ is the energy difference between the resonant level $\ket{k=0}$ and an off-resonant level $\ket{k}$.

When $g/ \sqrt{L} \ll \Delta_k$ and $\mu=0$, up to zeroth order, tunneling from $\ket{X}$ to $\ket{Y}$ mediated by $\ket{k=0}$ takes time $T = \pi \sqrt{L}/(\sqrt{2} g) = \pi / \Omega$. As in Sec.~\ref{sec:transfer_infidelity}, we apply perturbation theory to calculate the effect of off-resonant levels~\cite{Yao2013Quantum}. The leading term for the transfer infidelity $\epsilon = 1 - |\bra{Y} e^{-iHt} \ket{X}|^2$ gives
\begin{equation}
\epsilon \approx \Omega^2 \sum_{k \neq 0} \frac{1 + (-1)^k \cos(\Delta_k T)}{\Delta_k^2}, 
\label{Eq:Infidelity_perturbation_SM}
\end{equation}
when chemical potential $\mu =  \Omega^2 \sum_{k\neq 0} [1 - 3(-1)^k] / 2 \Delta_k$ is chosen to compensate for the level repulsion caused by off-resonant levels.

\begin{figure}
\hspace*{0\textwidth}
\includegraphics[width=0.63\linewidth]{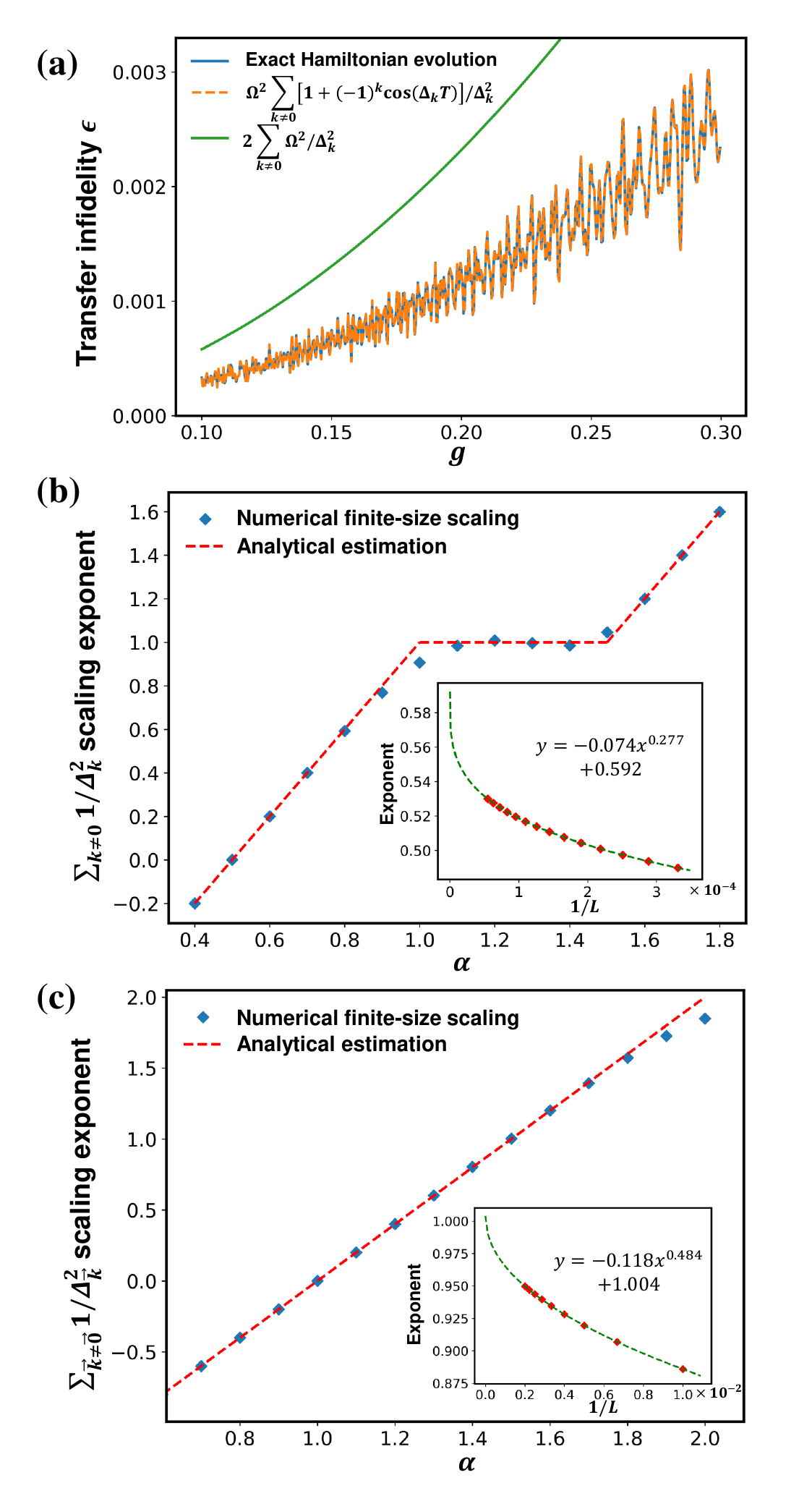} 
\caption{Numerical results on the state-transfer infidelity and state-transfer time of the time-independent protocol with $J_{ij}$ = $J_0/r_{ij}^\alpha$. (a) The state-transfer infidelity $\epsilon$ as a function of the coupling strength $g$ with $d=1$, $L=100$, and $\alpha=1$. (b) The extrapolated scaling exponents of the quantity $ \sum_{k \neq 0} 1 / \Delta_k^2 $ for different $\alpha$ in $d=1$, which agree well with our analytical estimation. Inset: Finite-size scaling of the locally computed exponents for $\alpha = 0.8$. We fit the red data points to the function $y = a x^b + c $ and obtain the intercept. (c) The extrapolated scaling exponents of the quantity $\sum_{k \neq 0} 1 / \Delta_k^2 $ for $\alpha \leq 2$ in $d=2$. Inset: Finite-size scaling of the locally computed exponents for $\alpha = 1.5$.}
\label{fig:Fig_S2}
\end{figure}

To benchmark the accuracy of the perturbation theory calculations, we numerically perform Hamiltonian evolution under Eq.~\eqref{Eq:hopping_Hamiltonian}
from initial state $\ket{X}$ for time $T = \pi / \Omega$ and compute the final state fidelity with respect to $\ket{Y}$. As illustrated in Fig.~\ref{fig:Fig_S2}(a), the resulting state-transfer infidelity is captured very well by the perturbation theory calculations Eq.~\eqref{Eq:Infidelity_perturbation_SM}.
From Fig.~\ref{fig:Fig_S2}(a) and Eq.~\eqref{Eq:Infidelity_perturbation_SM}, we observe that, when decreasing the coupling strength $g$, the transfer infidelity is upper bounded by $2 \sum_{k \neq 0} \Omega^2 / \Delta_k^2$. Therefore, for a given constant allowed infidelity $\epsilon$, the quantum state-transfer time scales as 
\begin{equation}
T = \mathcal{O}\left( \sqrt{ \frac{1}{\epsilon} \sum_{k \neq 0} \frac{1}{\Delta_k^2} } \right).
\label{Eq:scaling_time_vs_eps_Delta_SM}
\end{equation}

We numerically calculate the quantity $ \sum_{k \neq 0} 1 / \Delta_k^2 $ for all $\alpha \ge 0$ using system sizes up to about $10^5$, and extract the local power-law scaling exponents by fitting the data within sliding intervals (of length $10^2$) of $L$. As shown in the inset of Fig.~\ref{fig:Fig_S2}(b), we carry out finite-size scaling for the locally computed scaling exponents to extrapolate the values in the thermodynamic limit. Specifically, we fit the locally computed scaling exponents versus $1/L$ to the function $y = a x^b + c $. The intercepts give the blue data points shown in Fig.~\ref{fig:Fig_S2}(b), which suggest the following scaling of the quantum state-transfer times:
\begin{equation}
T = 
\begin{cases}
\mathcal{O}(L^{\alpha - 1/2}), & \alpha < 1 \\
\mathcal{O}( \sqrt{L} ), & 1 \leq \alpha < 3/2\\
\mathcal{O}(L^{\alpha - 1}), & 3/2 \leq \alpha < 2 \\
\mathcal{O}( L ), & \alpha \geq 2.
\end{cases}
\label{Eq:Transfer_time_1D}
\end{equation}


\begin{figure}
\hspace*{0\textwidth}
\includegraphics[width=.92\linewidth]{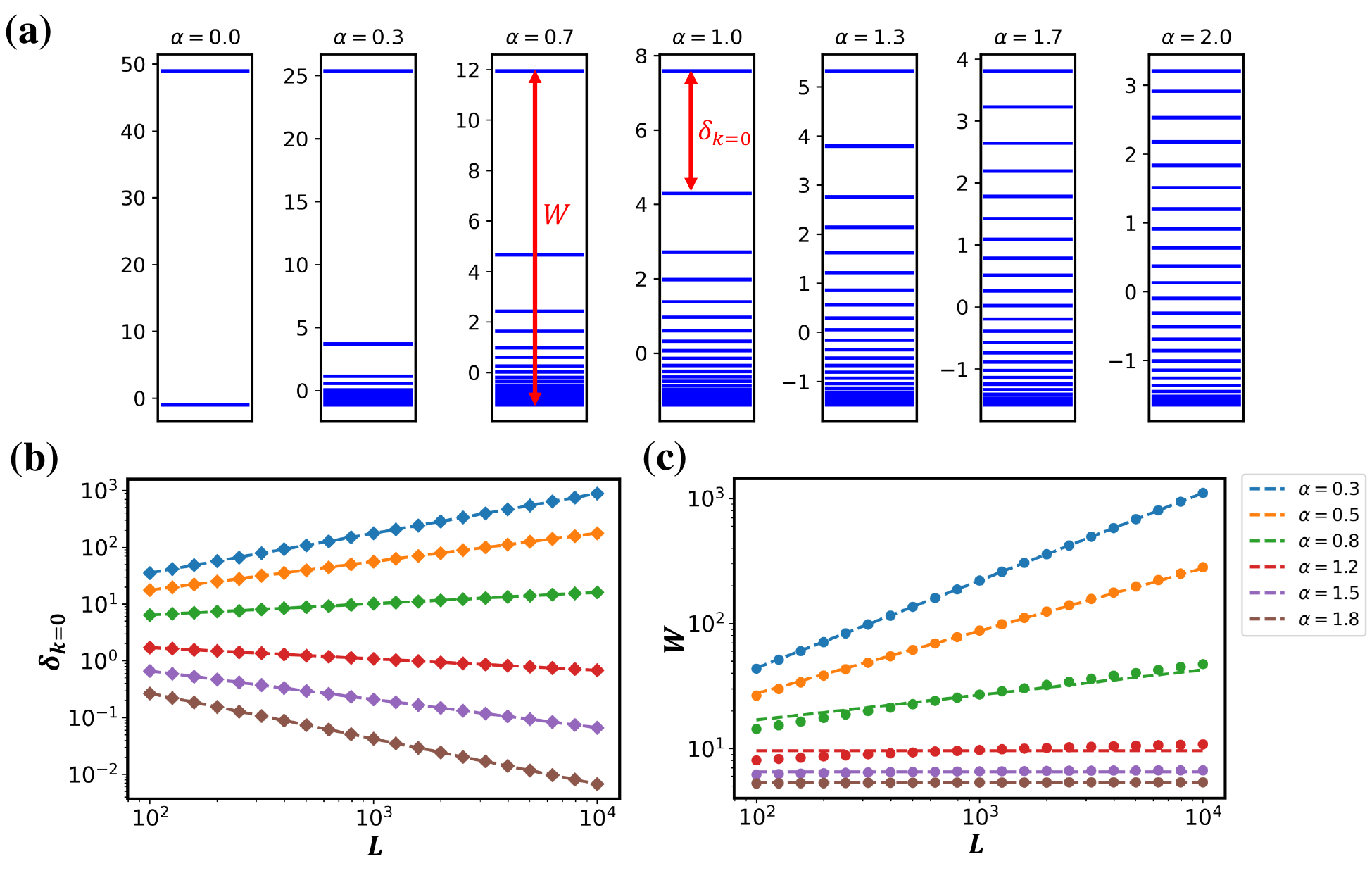} 
\caption{Numerical results regarding the single-particle spectrum properties for $d=1$. (a) The single-particle spectrum $\{E_k\}$ for different $\alpha$ with $L=50$. (b) The scaling behavior of the energy gap $\delta_{k=0}$ with respect to the system size $L$. The dashed lines are fits to the function $y = (1-\alpha) x + c$. (c) The scaling behavior of the spectrum bandwidth $W$ with respect to $L$. The dashed lines are fits to the function $y = (\max\{1-\alpha, 0\}) x + c$.}
\label{fig:Fig_S3}
\end{figure}

Besides the numerical results, we also provide a simple heuristic analytical argument to explain the scaling behavior in Eq.~\eqref{Eq:Transfer_time_1D}.
As depicted in Fig.~\ref{fig:Fig_S3}, the evaluation of $ \sum_{k \neq 0} 1 / \Delta_k^2 $ mainly involves two quantities characterizing the single-particle spectrum [in Fig.~\ref{fig:Fig_S3}(a), we display the spectrum for several different $\alpha$]. The first quantity  is the gap between the mode used for state transfer and adjacent energy levels. As shown in Fig.~\ref{fig:Fig_S3}(a), we numerically find that, for $\alpha \leq 2$, the largest energy gap always appears near the momentum eigenmode $\ket{k = 0}$, which explains why we choose $\ket{k = 0}$ as the resonant level for tunneling. 
In Fig.~\ref{fig:Fig_S3}(b), we demonstrate that the energy gap $\delta_{k=0}$ scales as $\delta_{k=0} = \Theta(L^{1-\alpha})$ for $\alpha \leq 2$.
The second quantity is the bandwidth $W$ of the whole single-particle spectrum, which, as displayed in Fig.~\ref{fig:Fig_S3}(c), scales as $W=\Theta(L^{1-\alpha})$ for $\alpha<1$, $W=\Theta( \log L )$ for $\alpha=1$, and $W = \Theta(1)$ for $\alpha>1$. The scaling behavior of $\delta_{k=0}$ and $W$ can also be obtained through analytical derivations presented in Eq.~\eqref{Eq:delta_k0_scaling} and Eq.~\eqref{Eq:W_scaling} below.

For $\alpha < 1$, since $\delta_{k=0}$ and $W$ are both $\Theta(L^{1-\alpha})$, we deduce that 
\begin{equation}
\sum_{k \neq 0}  \frac{1}{W^2} \leq \sum_{k \neq 0} \frac{1}{\Delta_k^2} \leq \sum_{k \neq 0} \frac{1}{ (\delta_{k=0})^2 }, 
\label{Eq:Scaling_alpha_l1}
\end{equation}
such that $\sum_{k \neq 0} 1 / \Delta_k^2 = \mathcal{O}( L \times L^{2(\alpha-1)} ) = \mathcal{O}( L^{2 \alpha - 1})$.

When $\alpha > 1$, inspired by numerics, we approximate the spectrum as a number of levels evenly spaced over the band, with the remainder piled up at the bottom of the spectrum. In this model, there can be about $W / \delta_{k=0} = \mathcal{O}(L^{\alpha - 1})$ levels evenly spaced between the top and bottom of the spectrum. Meanwhile, $\mathcal{O}(L - L^{\alpha - 1} )$ other levels pile up near the bottom of the spectrum. Therefore, the quantity $\sum_{k \neq 0} 1 / \Delta_k^2$ can be estimated as 
\begin{equation}
\mathcal{O} \left[ \frac{1}{ (L^{1-\alpha})^2 } + \frac{1}{( 2 L^{1-\alpha})^2} + \cdots + \frac{1}{  (L^{\alpha-1} L^{1-\alpha})^2 } + (L - L^{\alpha - 1}) \times \frac{1}{ W^2 }  \right] 
= \mathcal{O}( L^{2(\alpha - 1)} ) + \mathcal{O}( L ).
\label{Eq:Scaling_alpha_g1}    
\end{equation}


We observe that there are two competing terms in Eq.~\eqref{Eq:Scaling_alpha_g1}, and the turning point is $\alpha = 3/2 $. When $1 < \alpha < 3/2 $, we have $\sum_{k \neq 0} 1 / \Delta_k^2 = \mathcal{O}( L )$, whereas when $3/2 \leq \alpha < 2$, we have $\sum_{k \neq 0} 1 / \Delta_k^2 = \mathcal{O}( L^{2(\alpha-1)} )$.
Therefore, according to Eq.~\eqref{Eq:scaling_time_vs_eps_Delta_SM}, this analytical argument successfully illustrates our numerical results regarding the state-transfer time scaling in Fig.~\ref{fig:Fig_S2}(b).


\subsection{ The \texorpdfstring{$d\ge 2$}{d≥2} cases }

In this subsection, we generalize the time-independent protocol with $J_{ij}$ = $J_0/r_{ij}^\alpha$ to higher spatial dimensions. Consider the single-particle long-range hopping Hamiltonian on a $d$-dimensional cubic lattice with periodic boundary conditions:
\begin{equation}
H = \sum_{\vec{x}, \vec{y}\neq \vec{x}  }  J_{\vec{x}, \vec{y}}\  c^\dagger_{\vec{x}}\  c_{\vec{y}}  + g \left[ c_X^\dagger c_{(0,0,\cdots,0)} + c_{ (\frac{L}{2},\frac{L}{2},\ldots,\frac{L}{2}) }^\dagger c_{Y} + \hc \right]  + \mu \left(c_X^\dagger c_X + c_{Y}^\dagger c_{Y} \right).
\label{Eq:higher_d_hopping_Ham}   
\end{equation}
Here $\vec{x} = (x_1, x_2, \ldots, x_d)$ denotes the spatial coordinates of a site. The coupling strength between sites $\vec x$ and $\vec y$ is $J_{\vec{x}, \vec{y}} = J_0 / r_{ \vec{x}, \vec{y} }^{\alpha}$, where $r_{ \vec{x}, \vec{y} } = \sqrt{ \sum_{i=1}^d (\min\{|x_i-y_i|, L - |x_i-y_i|\})^2 }$ and we set $J_0=1$.
We use $L$ to denote the linear system size and $N = L^d$ to denote the total number of sites. The two external vertices $X$ and $Y$ are attached to sites $(0,0,\cdots,0)$ and $(L/2,L/2,\cdots,L/2)$, respectively. 
Through Fourier transformation of the first term in Eq.~\eqref{Eq:higher_d_hopping_Ham},
\begin{equation}
c_{\vec{k}} = \frac{1}{\sqrt{N}}\sum_{ \vec{x} } e^{ i 2 \pi \vec{k} \cdot \vec{x} / L } c_{\vec{x}}, \quad \vec{k} = (k_1, k_2 \cdots, k_d), \quad k_i = 0,1,\cdots,L-1,
\end{equation}
we obtain the single-particle spectrum of the quantum state-transfer channel:
\begin{equation}
E_{\vec{k}} = 2^d \sum_{x_1,x_2,\cdots,x_d = 1}^{L/2-1} \frac{ \prod_{i=1}^d \cos(2 \pi k_i x_i / L)}{(\sum_{i=1}^d x_i^2)^{\alpha / 2}} + \text{lower-dimension contributions}, \qquad k_i = 0,1,\cdots,L-1.
\label{Eq:spectrum_higher_d}
\end{equation}
The lower-dimension contributions in Eq.~\eqref{Eq:spectrum_higher_d} correspond to cases where certain coordinates are fixed to be zero and the single particle only hops on a lower-dimensional manifold embedded within the $d$-dimensional lattice. For example, when $d=2$, the single-particle spectrum reads (omitting the unimportant $x, y = L/2$ terms since they do not affect the scaling behaviors)
\begin{equation} 
E_{(k_x,k_y)} = 4 \sum_{x,y=1}^{L/2 - 1}\frac{\cos( 2 \pi k_x x / L) \cos( 2\pi k_y y / L ) }{ (x^2 + y^2)^{\alpha/2} } + 2 \sum_{x=1}^{L/2 - 1} \frac{\cos( 2 \pi k_x x / L)}{x^\alpha}  + 2 \sum_{y=1}^{L/2 - 1} \frac{\cos( 2\pi k_y y / L )}{ y^\alpha} .
\label{Eq:2d_spectrum}
\end{equation}


Similarly to the $d=1$ case, we resonantly couple states $\ket{X}$ and $\ket{Y}$ to the zero-momentum eigenmode $\ket{\vec{k}=\vec{0}}$ to carry out the quantum state transfer. By shifting the energy of $\ket{\vec{k}=\vec{0}}$ to be zero, we rewrite the Hamiltonian Eq.~\eqref{Eq:higher_d_hopping_Ham} as
\begin{equation}
H = \sum_{\vec{k}} \Delta_{\vec{k}}\  c_{\vec{k}}^\dagger \  c_{\vec{k}}  + \frac{g}{\sqrt{N}} \sum_{\vec{k}}  \left[ c_X^\dagger c_{\vec{k}} + (-1)^{\sum_{i=1}^d k_i} c_{\vec{k}}^\dagger c_{Y} + \hc \right] + \mu (c_X^\dagger c_X + c_{Y}^\dagger c_{Y}),  
\label{Eq:Fourier_Ham_dg2}
\end{equation}
where $\Delta_k = E_{\vec{k}=\vec{0}} - E_{\vec{k}}$. By noticing the similarity between Eq.~\eqref{Eq:Fourier_Ham_d=1} and Eq.~\eqref{Eq:Fourier_Ham_dg2}, we deduce that, after the evolution time $T=\pi \sqrt{N} / (\sqrt{2} g) = \pi/ \Omega$, the state-transfer infidelity becomes  
\begin{equation}
\epsilon \approx \sum_{\vec{k} \neq \vec{0}} \left(\frac{\Omega}{\Delta_{\vec{k}}}\right)^2 \left[1 + (-1)^{\sum_{i=1}^d k_i} \cos \Delta_{\vec{k}} T \right].
\label{Eq:infidelity_higher_dim}
\end{equation}
Hence, the scaling behavior of the quantum state-transfer time $T$ is still controlled by the quantity $\sum_{\vec{k} \neq \vec{0}} 1 / \Delta_{\vec{k}}^2 $.

 
As illustrated previously, the evaluation of $\sum_{\vec{k} \neq \vec{0}} 1 / \Delta_{\vec{k}}^2 $ mainly involves the largest energy gap $\delta_{\vec{k}=\vec{0}}$ and the single-particle spectrum bandwidth $W$. The scaling behavior of both of these quantities can be derived based on the leading term in Eq.~\eqref{Eq:spectrum_higher_d}:
\begin{equation}
\delta_{\vec{k}=\vec{0}} 
\propto \sum_{\forall i, x_i = 1}^{L/2-1} \frac{1- \prod_{i=1}^d \cos(2 \pi k_i x_i / L)}{(\sum_{i=1}^d x_i^2)^{\alpha / 2}} 
\propto \frac{1}{L^2} \sum_{\forall i, x_i = 1}^{ \mathcal{O}(L) } \frac{ \sum_{i=1}^d x_i^2 }{(\sum_{i=1}^d x_i^2)^{\alpha / 2}} 
\propto \frac{1}{L^2}\int_{\mathcal{O}(1)}^{\mathcal{O}(L) } \frac{r^2}{r^\alpha} r^{d-1} d r = \Theta(L^{d-\alpha}),   
\label{Eq:delta_k0_scaling}
\end{equation}
and
\begin{equation}
W \propto \sum_{\forall i, x_i = 1}^{L/2-1} \frac{1}{(\sum_{i=1}^d x_i^2)^{\alpha / 2}} 
\propto \int_{\mathcal{O}(1)}^{\mathcal{O}(L) } \frac{1}{r^\alpha} r^{d-1} d r = 
\begin{cases}
\Theta(L^{d - \alpha }), & \alpha < d \\
\Theta( \log L ), & \alpha = d\\
\Theta(1), & \alpha >d.
\end{cases}
\label{Eq:W_scaling}
\end{equation}
In the second step of Eq.~\eqref{Eq:delta_k0_scaling}, we set one $k_i=1$ and other $k_{j\neq i}=0$.


Using an inequality similar to the one in Eq.~\eqref{Eq:Scaling_alpha_l1}, when $\alpha < d$, we obtain
\begin{equation}
\sum_{\vec{k} \neq \vec{0}}  \frac{1}{\Delta_{\vec{k}}^2} = \mathcal{O}\left( L^d \times \frac{1}{ L^{2(d-\alpha)} } \right) = \mathcal{O}( L^{2\alpha - d} ),
\label{Eq:Scaling_alpha_ld_higher_d}
\end{equation}
which directly gives the quantum state-transfer time $T = \mathcal{O}( L^{\alpha - d/2} )$ for $\alpha < d$. As shown in Fig.~\ref{fig:Fig_S2}(c), we have carried out the finite-size scaling analysis to numerically confirm the result Eq.~\eqref{Eq:Scaling_alpha_ld_higher_d}.

One essential and different point from the $d=1$ case is that when $\alpha$ increases to $d/2 + 1$, the $d\ge2$ protocol has already reached the linear state-transfer time $T = \mathcal{O}( L )$. However, the linear transfer time can be easily achieved by only utilizing a one-dimensional chain with nearest-neighbor hopping  on the $d$-dimensional lattice. This observation means that, when $d \geq 2$ and $\alpha > d/2 + 1$, the current protocol with $J_{ij}=J_0/r_{ij}^\alpha$  provides no advantage over one-dimensional nearest-neighbor hopping. 
This phenomenon likely arises from the increased energy degeneracy (or near-degeneracy) appearing in dimensions $d\ge 2$, which also causes the estimation in
Eq.~\eqref{Eq:Scaling_alpha_g1} to break down when $d\ge 2$ and $\alpha \ge d$.



When compared with the free-particle Lieb-Robinson bounds, as pictorially summarized in Fig.~\ref{fig:Fig_S1}(c) and (d),  our time-independent protocol with $J_{ij}=J_0/r_{ij}^\alpha$ is optimal for $d = 1$ when $\alpha \leq 1/2$ or $\alpha\ge 3/2$ and for $d \ge 2$ when $\alpha \leq d/2 $. Despite overall being suboptimal, this time-independent translation-invariant protocol still provides a speed-up over short-range protocols for certain regimes of $\alpha$, showing the potential of using time-independent long-range Hamiltonians to accelerate quantum information processing tasks.




\end{document}